\documentclass[twocolumn]{aastex62}
\usepackage{xcolor}
\usepackage[utf8]{inputenc}
\received{}
\revised{}
\accepted{}

\shorttitle{Type Ib/Ic supernovae: effect of nickel mixing}
\shortauthors{Yoon et al.}

\begin{document}

\title{Type I\lowercase{b}/I\lowercase{c} supernovae: 
effect of nickel mixing on the early-time color evolution
and implications for the progenitors}

\correspondingauthor{Sung-Chul Yoon}
\email{yoon@astro.snu.ac.kr}

\author{Sung-Chul Yoon}
\affiliation{Department of Physics and Astronomy, Seoul National University, 08826, Seoul, South Korea}
\affiliation{Center for Theoretical Physics (CTP), Seoul National University, 08826, Seoul, South Korea}
\affiliation{Monash Center for Astrophysics, Monash University, Australia}

\author{Wonseok Chun}
\affiliation{Department of Physics and Astronomy, Seoul National University, 08826, Seoul, South Korea}

\author{Alexey Tolstov}
\affiliation{Kavli Institute for the Physics and Mathematics of the Universe (WPI), The University of Tokyo Institutes for Advanced Study, The University of Tokyo, 5-1-5 Kashiwanoha, Kashiwa, Chiba 277-8583, Japan}

\author{Sergey Blinnikov}
\affiliation{Kavli Institute for the Physics and Mathematics of the Universe (WPI), The University of Tokyo Institutes for Advanced Study, The University of Tokyo, 5-1-5 Kashiwanoha, Kashiwa, Chiba 277-8583, Japan}
\affiliation{Institute for Theoretical and Experimental Physics (ITEP), 117218 Moscow, Russia}
\affiliation{All-Russia Research Institute of Automatics (VNIIA), 127055 Moscow, Russia}

\author{Luc Dessart}
\affiliation{Unidad Mixta Internacional Franco-Chilena de Astronom\'{i}a (CNRS UMI 3386), Departamento de Astronomía, Universidad de
Chile, Camino El Observatorio 1515, Las Condes, Santiago, Chile}



\begin{abstract}
We investigate the effect of mixing of radioactive nickel ($^{56}$Ni) on the
early-time color evolution of Type Ib and Ic supernovae (SNe Ib/Ic) using
multi-group radiation hydrodynamics simulations. 
We consider both helium-rich and helium-poor progenitors. 
Mixing of $^{56}$Ni is parameterized using
a Gaussian distribution function. We find that the early-time color evolution
with a weak $^{56}$Ni mixing is characterized by three different phases: initial rapid reddening, blueward
evolution due to the delayed effect of $^{56}$Ni heating, and redward evolution thereafter
until the transition to the nebular phase.  With a strong $^{56}$Ni
mixing, the second phase disappears.  We
compare our models with the early-time color evolution of several SNe Ib/Ic (SN
1999ex, SN 2008D, SN 2009jf, iPTF13bvn, SN 1994I, SN
2007gr, SN 2013ge, and 2017ein) and find signatures of  relatively weak and strong
$^{56}$Ni mixing  for SNe Ib and SNe Ic, respectively. This suggests
that  SNe Ib progenitors  are distinct from SN Ic progenitors
in terms of helium content and that $^{56}$Ni mixing 
is generally stronger in the carbon-oxygen core and weaker in the helium-rich envelope. 
We conclude that the early-time color evolution
is a powerful probe of $^{56}$Ni mixing in SNe Ib/Ic.   
\end{abstract}


\section{Introduction}\label{sect:Introduction}

Type Ib and Ic supernovae (SNe Ib/Ic) and their progenitors constitute an
important subject of stellar evolution theory \citep[see][for a recent
review]{Yoon2015}.  The majority of SNe Ib/Ic belong to a subclass of
core-collapse SNe (ccSNe), having a massive star origin as implied by their
strong correlation with star formation 
\citep[e.g.,][]{Tsvetkov2001, vandenBergh2005, Anderson2009, Hakobyan2009, Habergham2010,
Leloudas2011, Kelly2012, Sanders2012, Hakobyan2014, Anderson2015, Xiao2015,
Kangas2017, Graur2017, Maund2018}. Their event rate is relatively high,
amounting to about 25 \% of the ccSN rate \citep[e.g.,][]{Smith2011,
Eldridge2013, Shivvers2017}.  The lack of \ion{H}{1} spectral lines 
implies that the progenitors have lost their hydrogen-rich
envelope during the pre-SN evolution by stellar wind and/or binary
interaction \citep[e.g.,][]{Podsiadlowski1992, Woosley1993, Woosley1995,
Wellstein1999,Eldridge2004, Meynet2005, Eldridge2008, Yoon2010, Georgy2012,
Yoon2017a}.  The properties of SN Ib/Ic light curves and spectra imply
relatively low ejecta masses (i.e., $M_\mathrm{ejecta} \simeq 1.0 -
4.0~\mathrm{M_\odot}$; e.g., \citealt{Ensman1988, Shigeyama1990, Dessart2011,
Drout2011, Bianco2014, Taddia2015, Lyman2016, Prentice2016, Taddia2018}), in favor of the
binary scenario for their progenitors~\citep{Yoon2015}.  The optically bright
progenitor of the SN Ib iPTF13bvn, which has been recently identified in a
pre-SN image~\citep{Cao2013,Folatelli2016}, is also consistent with binary
progenitor models in terms of the progenitor mass and optical
brightness (\citealt{Bersten2014, Eldridge2015, Kim2015, Fremling2016,
McClelland2016, Hirai2017, Yoon2017a}; see also \citealt{Yoon2012}).

The exact nature of SN Ib/Ic progenitors is still a matter of debate.  One
outstanding question is what distinguishes type Ib from type Ic progenitors. 

Supernovae Ib/Ic are distinguished by the presence or absence of
\ion{He}{1} lines. Because of this reason,  SN Ic progenitors are often
regarded as helium-poor stars in the literature. However, the absence of
\ion{He}{1} lines in SN Ic spectra does not necessarily serve as evidence for
the deficiency of helium in SN Ic progenitors.  This is because the formation
of \ion{He}{1}  lines by non-thermal processes  \citep{Lucy1991,Swartz1991} is sensitively
affected by the presence of radioactive $^{56}$Ni, and hence by the degree of
$^{56}$Ni mixing in SN ejecta~\citep{Woosley1997, Dessart2012}.  \ion{He}{1} lines can
be easily formed if $^{56}$Ni were fully mixed in the SN ejecta but a large
amount of helium ($M_\mathrm{He} \gtrsim 1.0~M_\odot$) could be effectively
hidden in the spectra with a very weak $^{56}$Ni  mixing~\citep{Woosley1997,
Dessart2012, Hachinger2012}.  For an understanding of the nature of SN Ib/Ic
progenitors, therefore,  it would be very useful if the $^{56}$Ni distribution
in SN Ib/Ic ejecta could be constrained by observations.

The effects of $^{56}$Ni mixing on SN Ib/Ic light curves and spectra have been
studied by several authors~\citep{Ensman1988, Shigeyama1990, Woosley1997,
Dessart2012, Bersten2013, Piro2013, Dessart2015, Dessart2016}. It is found that
some degree of $^{56}$Ni mixing is needed not only for the efficient formation
of \ion{He}{1} lines in the spectra~\citep{Woosley1997, Dessart2012} but also for
explaining the overall shape of SN Ib light curves during the post-maximum
phase ~\citep{Shigeyama1990}.  The early-time light curve is also found to be
significantly affected by the $^{56}$Ni distribution.  For example, the
post-breakout plateau of a short duration (a few to several days) that is
commonly found in the models with a relatively weak $^{56}$Ni mixing 
disappears with a strong  $^{56}$Ni mixing~\citep{Ensman1988, Dessart2011, Dessart2012, 
Piro2013}. 

In addition, a larger  $^{56}$Ni content in the outermost layers would lead to
a higher temperature of the SN Ib/Ic photosphere at very early times,
resulting in a bluer color for a given progenitor structure~\citep{Piro2013,
Dessart2015} -- a similar effect results for an extended progenitor 
\citep{Nakar2014,Dessart2018}.  This implies that we can use this information to constrain the
degree of  $^{56}$Ni mixing in SNe Ib/Ic but this possibility has not been
properly explored yet.  In this study, we present new multi-color light curve
models of SN Ib/Ic calculated with the STELLA code and systematically
investigate the effects of $^{56}$Ni mixing on the early-time evolution of SNe
Ib/Ic.  We show that the early-time color evolution of SN Ib/Ic sensitively
depends on the $^{56}$Ni distribution, and can be a powerful probe of $^{56}$Ni
mixing in SN Ib/Ic. 

This paper is organized as follows. In Section~\ref{sec:assumption}, we present
our progenitor models, the numerical method, 
and the physical parameters adopted for SN calculations. In Section~\ref{sec:LC},
we present the multi-color light curves of our models. We discuss the effect of
$^{56}$Ni mixing on the early-time color evolution in Section~\ref{sec:color}.
We compare our models with several observed SNe Ib/Ic including SN 1994I, SN
1999ex, SN 2007gr, SN 2008D, SN 2009jf, iPTF13bvn,  SN 2013ge and 2013ein in
Section~\ref{sec:comp}. We discuss the implications of our results for SNe Ib/Ic
progenitors and  $^{56}$Ni mixing in Section~\ref{sec:implication}. We conclude
our study in Section~\ref{sec:conclusion}.

\section{Physical assumptions and numerical method}\label{sec:assumption}

\subsection{Input models}\label{sec:input}

\begin{table}
{\footnotesize
\begin{center}
\caption{Input model properties} 
\label{tab1}
\begin{tabular}{l | r@{\hspace{2mm}} r@{\hspace{2mm}} r@{\hspace{2mm}} r@{\hspace{2mm}} 
r@{\hspace{2mm}} r@{\hspace{2mm}} r@{\hspace{2mm}} r@{\hspace{2mm}}} 
\hline
 Name       &  $M_\mathrm{ej}$ &  $R$  & $M_\mathrm{CO}$ & $M_\mathrm{He, env}$ & $m_\mathrm{He}$ & $M_\mathrm{Fe}$ & $M_\mathrm{ext}$  \\
            &  [$M_\odot$] &  [$R_\odot$]  & [$M_\odot$] & [$M_\odot$] & [$M_\odot$] & [$M_\odot$] & [$M_\odot$]  \\
\hline
\hline
HE3.87 &     2.40           &  6.73            &  2.15           &  1.72           &  1.66           &  1.47  & 2.7(-5) \\ 
\hline
CO3.93 &      2.49           &  0.77            &  3.91           &  0.02           &  0.10           &  1.44  &  5.2(-4)                   \\
CO2.16 &      0.71           &  0.13            &  2.16           &  0.00           &  0.06           &  1.46  &  3.2(-4)                  \\
\hline
\end{tabular}
\tablecomments{$M_\mathrm{ej}$: ejecta mass (i.e., total mass minus the mass cut);
$M_\mathrm{CO}$: CO core mass; $M_\mathrm{He, env}$: helium-rich envelope mass; $m_\mathrm{He}$: integrated helium mass; 
$M_\mathrm{Fe}$: iron core mass, which also corresponds to the adopted mass cut for the piston-driven explosion in {\sc stella}; $M_\mathrm{ext}$: mass of the external buffer used to limit the maximum ejecta velocity (see Section~\ref{sec:assumption} for discussion). Numbers in parenthesis denote powers of ten.} 
\end{center}
}
\end{table}

\begin{figure}
\centering
\includegraphics[width=0.999\columnwidth]{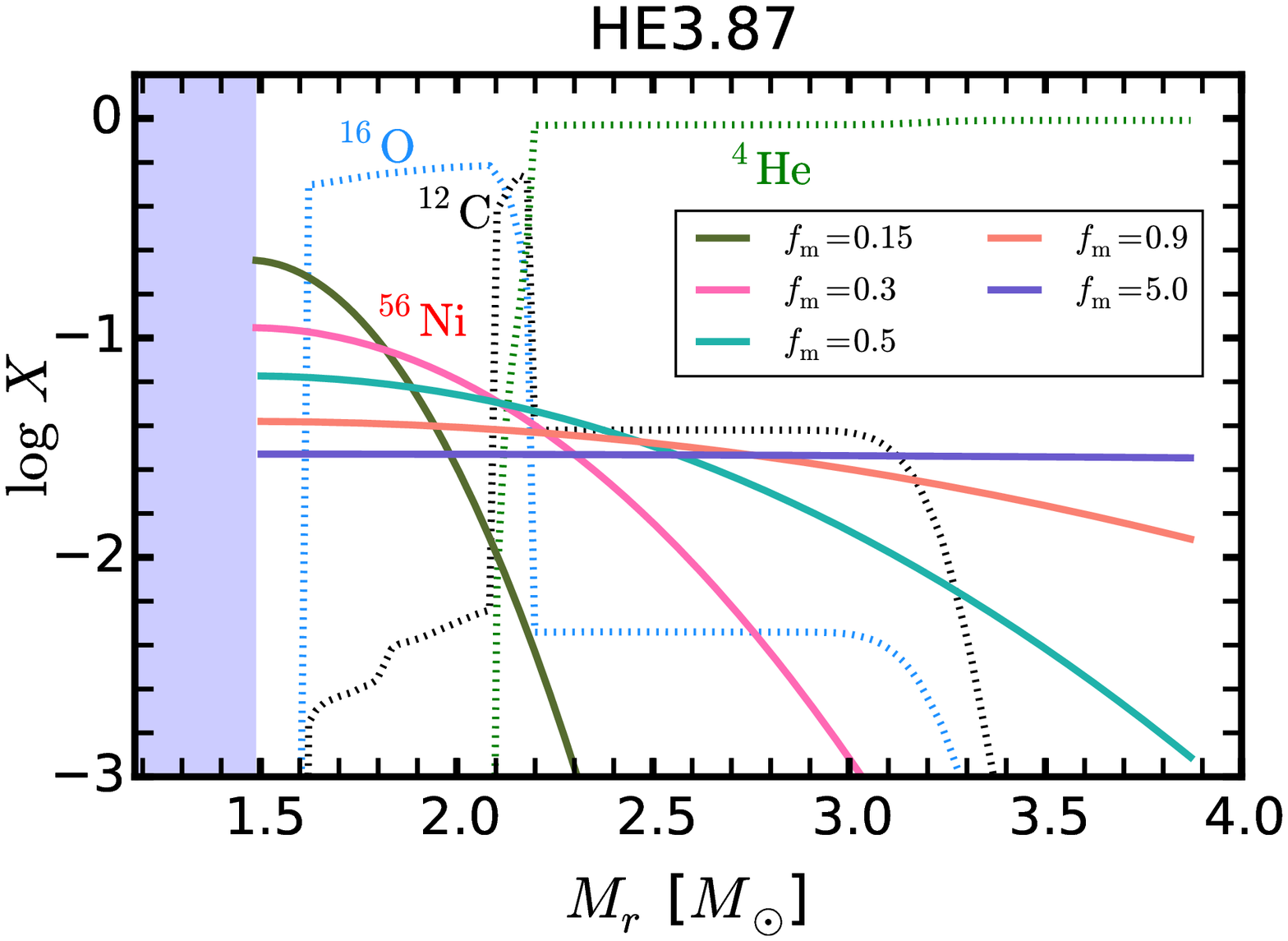}
\includegraphics[width=0.999\columnwidth]{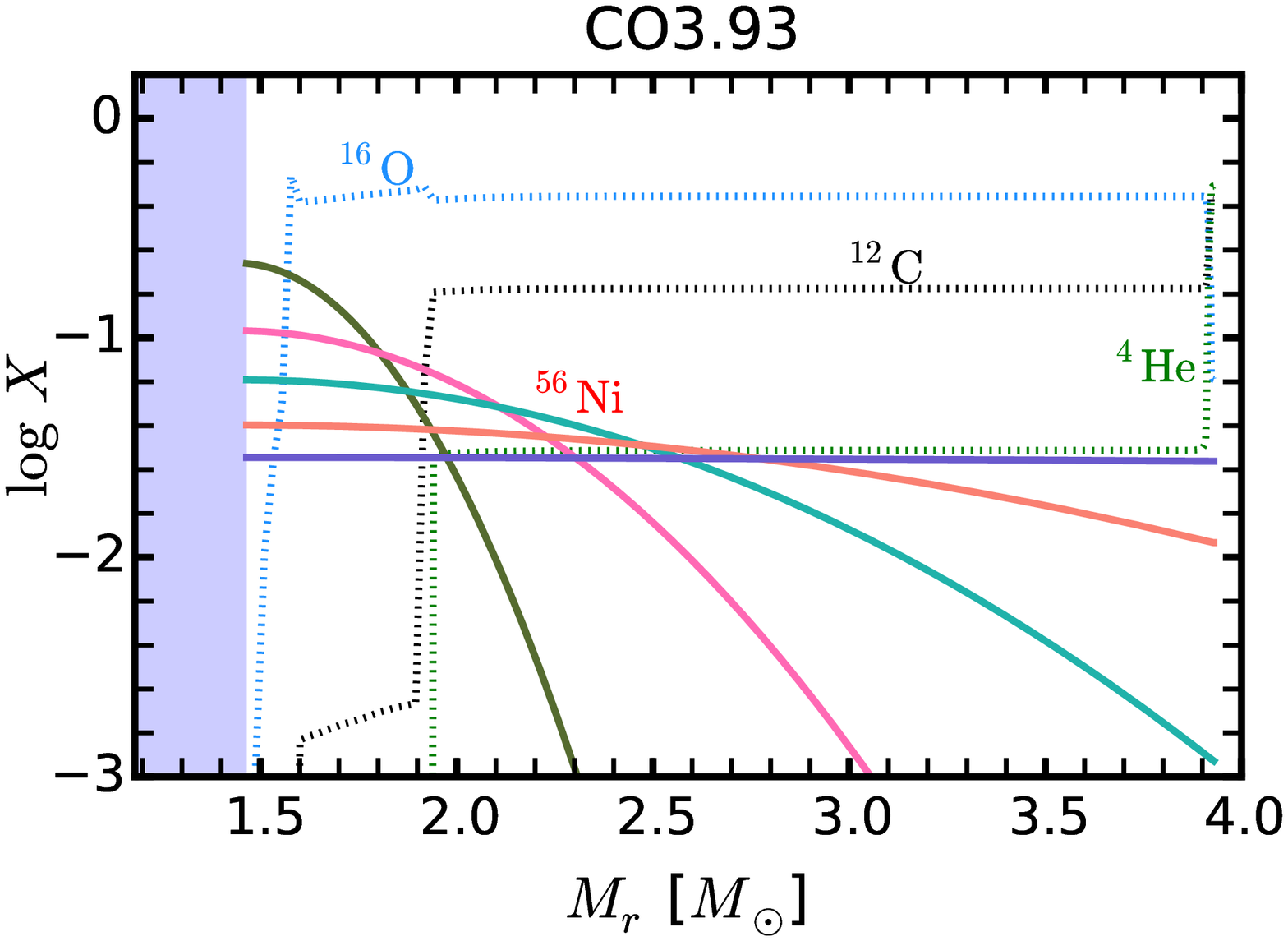}
\caption{The chemical composition in the input models HE3.87 (upper panel) and CO3.93 (lower panel).  
$^{56}$Ni profiles with $f_\mathrm{m} =$ 0.15, 0.3, 0.5, 0.9 and 5.0 according to equation~(\ref{eq1})
are color coded.} 
\label{fig:initial}
\end{figure}

\begin{figure}
\centering
\includegraphics[width=0.999\columnwidth]{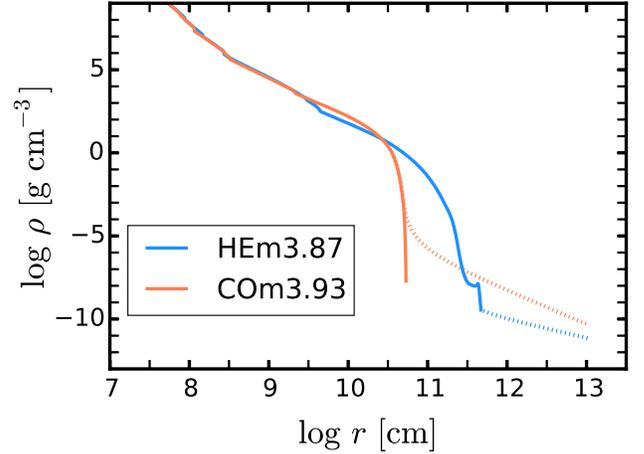}
\caption{The density profiles of the input models HE3.87 (orange) and CO3.93 (skyblue).   
The solid  and dotted lines give the density profiles of the progenitor models and 
the attached external material, respectively. See the text for details. 
}  
\label{fig:density}
\end{figure}

We consider both helium-rich and helium-poor progenitors.  These models are
constructed with the MESA code~\citep{Paxton2011, Paxton2013, Paxton2015,
Paxton2018}.  In Table~\ref{tab1}, we present the properties of our progenitor
models.  The boundary between the helium-rich envelope and the carbon-oxygen
(CO) core is defined by the condition that the mass fraction of helium in the
envelope is higher than 0.1.  The name starting with `HE' refers to the
helium-rich model, which has an integrated helium mass (i.e., $m_\mathrm{He} = \int
X_\mathrm{He} dM_r$) of about 1.7~$M_\odot$, and thus is a potential SN Ib
progenitor. Note that with this much helium,  \ion{He}{1} lines would be seen 
at early times even without non-thermal effects~\citep{Dessart2011}.  
The name with `CO' refers to the helium-poor model ($m_\mathrm{He} \le 
0.1 M_\odot$), which is  almost a naked CO core and can be considered a SN Ic
progenitor. The number given in each model name refers to the total mass of the
progenitor at the pre-SN stage. 

Model HE3.87 is from the binary star model sequence Sm13p50 of \citet{Yoon2017a},
where the primary star of 13.0~$M_\odot$ starts its evolution in a 50 day orbit
with a 11.7~$M_\odot$ companion.  The calculation of this model is stopped
when the central temperature reaches $10^9$~K in \citet{Yoon2017a}, and  the
evolution is  further followed up to the pre-SN stage for the present study.
Model CO2.16 has the same progenitor model and the same adopted physics 
as model HE3.87 but
the mass-loss rate is artificially enhanced (i.e., $\dot{M} = 10^{-4}~M_\odot
~\mathrm{yr^{-1}}$) after core helium exhaustion, in order to remove the helium-rich
envelope. 
Model CO3.93  is evolved from a pure helium star of 7.0~$M_\odot$ at solar
metallicity, which would have a $\sim$ 20~$M_\odot$ main sequence progenitor.   
Overshooting on top of the helium-burning core is
treated as a step function with an overshoot parameter of 0.1~$H_P$ ($H_P$ is
the pressure scale height at the boundary of the convective core).  The
standard mass-loss prescription for Wolf-Rayet (WR) stars by \citet{Nugis2000} is
adopted until core helium exhaustion  and an enhanced mass-loss rate of
$\dot{M} = 10^{-4}~M_\odot ~\mathrm{yr^{-1}}$ is applied thereafter until the
pre-SN stage to remove the helium-rich envelope. 

We assume that the mass cut in the ccSN explosion corresponds to the iron core
mass $M_\mathrm{Fe}$, which is the innermost region above which the silicon mass fraction 
is higher than 0.1.  
We put 0.07~$M_\odot$ of $^{56}$Ni by hand in the
progenitor model assuming a Gaussian distribution.  
The degree of $^{56}$Ni mixing is parameterized using the parameter $f_\mathrm{m}$, defined as 
\begin{equation}\label{eq1} 
X_\mathrm{Ni}(M_r) = A
\exp\left(-\left[\frac{M_r - M_\mathrm{Fe}}{f_\mathrm{m}(M_\mathrm{tot} -
M_\mathrm{Fe})}\right]^2\right)~, 
\end{equation} 
where $M_r$ is the Lagrangian mass at radius $r$, and $X_\mathrm{M_r}$
denotes the mass fraction of  $^{56}$Ni at $M_r$.
The total mass of $^{56}$Ni for a given
$f_\mathrm{m}$ is determined by the normalization factor $A$.  The mass
fractions of the other elements are scaled to preserve a normalization of the
total mass fraction to unity at each depth.
We consider $f_\mathrm{m} =$ 0.15, 0.3, 0.5, 0.9 and 5.0.  As shown in Figure
1, $^{56}$Ni is more evenly distributed with a larger value of $f_\mathrm{m}$.
Mixing of the other elements, which is not considered in this study, is
potentially important for the early-time light curves of helium-rich models
because mixing of carbon and oxygen into the helium-rich envelope can influence
the opacity at the SN photosphere. For example,  the importance of ionization
in the opacity would be reduced with more carbon and oxygen. This effect will
be addressed elsewhere. 

Our progenitor models are compact ($R = 6.73~R_\odot$, $0.77~R_\odot$ and
$0.13~R_\odot$ for HE3.87, CO3.93 and CO2.16, respectively; Table~\ref{tab1}),
and the density decreases sharply at the outermost layers
(Figure~\ref{fig:density}).  This would make the shock velocity rapidly
accelerate to a value close to the speed of light, while our version of the STELLA code 
does not include relativistic effects. To avoid
numerical errors that this high shock velocity can introduce, we add a small
amount of external material beyond the stellar surface 
(i.e., which also mimics the presence of an atmosphere or wind),  so that the velocity of the shock
front remains significantly below the speed of light.  This mass buffer is assumed to
have a wind density profile (i.e., $\rho = \dot{M}/4\pi v_\mathrm{w} r^2$) for
which the standard $\beta$-law for the wind velocity $v_\mathrm{w}$ with $\beta = 2.0$, $v_\infty =
10~\mathrm{km~s^{-1}}$ and $\dot{M} = 10^{-3}~M_\odot~\mathrm{yr^{-1}}$ is
assumed.  The maximum radius of this external material is set to $10^{13}$~cm initially
(Figure~\ref{fig:density}).  Because CO3.83 and CO2.16 are much more compact, the
density decrease at the outer boundary is too sharp and causes too small a time
step before shock breakout.  Therefore we remove 
$\sim 10^{-6}~M_\odot$ from the original progenitor model and attach the mass buffer as
shown in Figure~\ref{fig:density}.  With this configuration, the ejecta
velocity at the outer boundary remains lower than about
$4\times10^{4}~\mathrm{km~s^{-1}}$ in all of our SN models.  Although this external structure is 
unrealistic, its mass is very small (i.e., $M_\mathrm{ext}
\lesssim  5 \times 10^{-4} M_\odot$; Table~\ref{tab1}) and it affects the SN
light curves only for a short period ($ < 1$ d) after shock breakout.

\subsection{Supernova Models}\label{sec:snmodel}

\begin{table*}
\begin{center}
\caption{Properties of bolometric light curves of the SN models with $M_\mathrm{Ni} = 0.07 M_\odot$.}\label{tab2} 
\begin{tabular}{l | c c r r r r  } 
\hline
Name & $E_\mathrm{K}$   &  $L_\mathrm{Bol, max}$              & $t_\mathrm{Bol,{max}}$ & $\delta t_\mathrm{Bol, -1/2}$ & $\delta t_\mathrm{Bol, +1/2}$ & $M_\mathrm{Bol, max}$  \\
     &  [$10^{51}$ erg]   &  [$10^{42}~\mathrm{erg~s^{-1}}$]  &      [d]       &     [d]      &      [d]           &       [mag]              \\     
\hline
\hline
HE3.87\_fm0.15\_E1.0  & 0.67  &  1.69   &   23.82 &  9.56 &  14.91 & -16.85 \\
HE3.87\_fm0.3\_E1.0  & 0.67  &  1.67   &   23.08 &  9.13 &  15.86 & -16.84  \\
HE3.87\_fm0.5\_E1.0  & 0.69  &  1.64   &   22.11 &  11.02 &  14.21 & -16.82 \\
HE3.87\_fm0.9\_E1.0  & 0.69  &  1.62   &   21.14 &  13.31 &  12.70 & -16.81 \\
HE3.87\_fm5.0\_E1.0  & 0.69  &  1.61   &   18.08 &  12.21 &  14.19 & -16.80 \\
\hline
HE3.87\_fm0.15\_E1.5  & 1.17  &  1.80   &   19.41 &  6.98 &  15.50 & -16.92\\
HE3.87\_fm0.3\_E1.5  & 1.17  &  1.81   &   18.94 &  7.47 &  14.18 & -16.93 \\
HE3.87\_fm0.5\_E1.5  & 1.18  &  1.83   &   18.00 &  8.57 &  13.18 & -16.94 \\
HE3.87\_fm0.9\_E1.5  & 1.19  &  1.82   &   17.33 &  10.12 &  11.35 & -16.93 \\
HE3.87\_fm5.0\_E1.5  & 1.19  &  1.77   &   14.94 &  9.80 &  12.32 & -16.90  \\
\hline
HE3.87\_fm0.15\_E1.8  & 1.47  &  1.85   &   18.07 &  6.83 &  15.03 & -16.95\\
HE3.87\_fm0.3\_E1.8  & 1.47  &  1.85   &   17.99 &  7.76 &  14.75 & -16.95 \\
HE3.87\_fm0.5\_E1.8  & 1.48  &  1.89   &   17.51 &  8.82 &  12.07 & -16.98  \\
HE3.87\_fm0.9\_E1.8  & 1.49  &  1.85   &   15.00 &  8.65 &  12.51 & -16.95  \\
HE3.87\_fm5.0\_E1.8  & 1.49  &  1.84   &   15.29 &  10.47 &  9.96 & -16.94  \\
\hline
CO3.93\_fm0.15\_E1.0  & 0.57  &  1.64   &   30.85 &  11.42 &  15.34 & -16.82 \\
CO3.93\_fm0.3\_E1.0  & 0.58  &  1.53   &   29.19 &  13.16 &  16.96 & -16.75  \\
CO3.93\_fm0.5\_E1.0  & 0.58  &  1.45   &   26.02 &  13.51 &  20.20 & -16.69  \\
CO3.93\_fm0.9\_E1.0  & 0.58  &  1.40   &   23.16 &  15.06 &  23.04 & -16.65  \\
CO3.93\_fm5.0\_E1.0  & 0.58  &  1.36   &   17.44 &  11.74 &  27.88 & -16.62  \\
\hline
CO3.93\_fm0.15\_E1.5  & 1.07  &  1.87   &   24.93 &  9.74 &  19.73 & -16.97  \\
CO3.93\_fm0.3\_E1.5  & 1.08  &  1.79   &   24.86 &  12.08 &  18.85 & -16.91  \\
CO3.93\_fm0.5\_E1.5  & 1.08  &  1.70   &   22.10 &  11.04 &  19.43 & -16.86  \\
CO3.93\_fm0.9\_E1.5  & 1.08  &  1.62   &   18.97 &  11.85 &  18.99 & -16.81  \\
CO3.93\_fm5.0\_E1.5  & 1.08  &  1.55   &   15.32 &  10.25 &  21.27 & -16.76  \\
\hline
CO3.93\_fm0.15\_E1.8  & 1.37  &  1.99   &   23.20 &  9.70 &  18.49 & -17.03  \\
CO3.93\_fm0.3\_E1.8  & 1.38  &  1.88   &   21.52 &  8.98 &  18.50 & -16.97  \\
CO3.93\_fm0.5\_E1.8  & 1.38  &  1.78   &   21.23 &  11.18 &  16.96 & -16.91  \\
CO3.93\_fm0.9\_E1.8  & 1.38  &  1.70   &   17.91 &  11.25 &  16.80 & -16.86  \\
CO3.93\_fm5.0\_E1.8  & 1.38  &  1.61   &   15.04 &  10.22 &  18.80 & -16.80  \\
\hline
CO2.12\_fm0.15\_E1.0  & 0.60  &  2.20   &   11.78 &  5.62 &  12.91 & -17.14  \\
CO2.12\_fm0.3\_E1.0  & 0.60  &  2.27   &   11.44 &  5.81 &  14.09 & -17.18  \\
CO2.12\_fm0.5\_E1.0  & 0.60  &  2.20   &   12.07 &  6.22 &  12.23 & -17.14  \\
CO2.12\_fm0.9\_E1.0  & 0.60  &  2.10   &   11.00 &  6.36 &  12.03 & -17.09  \\
CO2.12\_fm5.0\_E1.0  & 0.60  &  2.03   &   10.05 &  6.35 &  11.49 & -17.05  \\
\hline
\end{tabular}
\tablecomments{$E_\mathrm{K}$: kinetic energy of SN,  $L_\mathrm{Bol, max}$: bolometric luminosity at the main peak,
$t_\mathrm{Bol,{max}}$: time from the shock breakout to the main peak of the bolometric luminsosity, 
$\delta t_\mathrm{Bol, -1/2}$: time span from $L_\mathrm{Bol, max}/2$ to $L_\mathrm{Bol, max}$, $\delta t_\mathrm{Bol, -1/2}$: time span  from  $L_\mathrm{Bol, max}$ 
to  $L_\mathrm{Bol, max}/2$,  $M_\mathrm{Bol, max}$: bolometric magnitude at the main peak.}
\end{center}
\end{table*}

\begin{table*}
{\footnotesize
\begin{center}
\caption{Properties of $U, B, V, R$-band light curves of SN models with $M_\mathrm{Ni} = 0.07 M_\odot$.}\label{tab3} 
\begin{tabular}{l | r r r r r r r r r r r r}
\hline
Name & $U_\mathrm{max}$ & $t_{U_\mathrm{max}}$ & $\delta t_{U,-1/2}$ & $B_\mathrm{max}$ & $t_{B_\mathrm{max}}$ & $\delta t_{B,-1/2}$ & $V_\mathrm{max}$ & $t_{V_\mathrm{max}}$ & $\delta t_{V,-1/2}$ & 
$R_\mathrm{max}$ & $t_{R_\mathrm{max}}$ & $\delta t_{R,-1/2}$  \\
\hline
\hline
HE3.87\_fm0.15\_E1.0  & -16.71  &  21.79   &   6.42 & -16.64 &  22.79 &  7.42 & -17.08 &   24.55 &  9.18 & -17.27 &   24.55 &  10.29 \\
HE3.87\_fm0.3\_E1.0  & -16.43  &  21.37   &   7.42 & -16.53 &  21.37 &  7.42 & -17.10 &   23.08 &  9.13 & -17.33 &   24.33 &  10.38 \\
HE3.87\_fm0.5\_E1.0  & -15.94  &  19.13   &   9.10 & -16.36 &  20.73 &  10.69 & -17.12 &   21.45 &  10.36 & -17.41 &   23.28 &  11.09 \\
HE3.87\_fm0.9\_E1.0  & -15.77  &  10.98   &   6.87 & -16.23 &  17.70 &  11.92 & -17.12 &   18.58 &  10.75 & -17.45 &   21.61 &  12.70 \\
HE3.87\_fm5.0\_E1.0  & -16.03  &  9.18   &   6.11 & -16.28 &  11.20 &  6.79 & -17.13 &   18.08 &  11.83 & -17.45 &   19.94 &  12.27 \\
\hline
HE3.87\_fm0.15\_E1.5  & -16.89  &  18.59   &   6.16 & -16.68 &  18.59 &  6.16 & -17.12 &   19.41 &  6.98 & -17.32 &   20.44 &  8.01 \\
HE3.87\_fm0.3\_E1.5  & -16.60  &  16.89   &   5.42 & -16.61 &  18.32 &  6.85 & -17.18 &   19.54 &  8.07 & -17.42 &   20.66 &  9.19 \\
HE3.87\_fm0.5\_E1.5  & -16.11  &  16.49   &   7.98 & -16.49 &  17.05 &  8.07 & -17.23 &   18.00 &  8.57 & -17.50 &   19.61 &  10.18 \\
HE3.87\_fm0.9\_E1.5  & -15.91  &  11.20   &   8.16 & -16.37 &  13.82 &  8.79 & -17.24 &   16.91 &  9.70 & -17.55 &   17.71 &  9.48 \\
HE3.87\_fm5.0\_E1.5  & -16.10  &  7.43   &   4.36 & -16.41 &  11.87 &  8.28 & -17.23 &   14.94 &  9.58 & -17.54 &   16.92 &  10.54 \\
\hline
HE3.87\_fm0.15\_E1.8  & -16.99  &  16.66   &   5.41 & -16.69 &  17.31 &  6.07 & -17.12 &   18.65 &  7.40 & -17.33 &   19.17 &  7.92 \\
HE3.87\_fm0.3\_E1.8  & -16.70  &  15.59   &   4.89 & -16.64 &  16.78 &  6.55 & -17.19 &   18.68 &  7.99 & -17.45 &   20.43 &  9.74 \\
HE3.87\_fm0.5\_E1.8  & -16.20  &  14.12   &   6.04 & -16.54 &  15.77 &  7.68 & -17.26 &   17.08 &  8.39 & -17.55 &   18.37 &  9.16 \\
HE3.87\_fm0.9\_E1.8  & -16.00  &  10.17   &   7.08 & -16.45 &  14.06 &  9.27 & -17.27 &   15.00 &  8.65 & -17.55 &   16.63 &  9.77 \\
HE3.87\_fm5.0\_E1.8  & -16.16  &  7.03   &   3.94 & -16.47 &  11.08 &  7.70 & -17.26 &   12.89 &  7.87 & -17.56 &   15.29 &  9.28 \\
\hline
CO3.93\_fm0.15\_E1.0  & -15.81  &  27.65   &   7.21 & -16.23 &  28.69 &  9.26 & -16.92 &   30.85 &  11.42 & -17.26 &   30.85 &  11.42 \\
CO3.93\_fm0.3\_E1.0  & -15.46  &  24.31   &   8.27 & -16.07 &  26.36 &  10.32 & -16.89 &   29.19 &  13.16 & -17.25 &   30.01 &  12.96 \\
CO3.93\_fm0.5\_E1.0  & -14.84  &  19.72   &   9.92 & -15.79 &  22.76 &  12.28 & -16.85 &   26.02 &  14.53 & -17.26 &   26.02 &  13.51 \\
CO3.93\_fm0.9\_E1.0  & -15.00  &  10.45   &   6.47 & -15.72 &  13.54 &  8.28 & -16.80 &   20.59 &  14.09 & -17.25 &   23.16 &  15.06 \\
CO3.93\_fm5.0\_E1.0  & -16.11  &  9.19   &   5.88 & -16.24 &  10.56 &  5.89 & -16.87 &   15.06 &  9.14 & -17.18 &   21.45 &  14.56 \\
\hline
CO3.93\_fm0.15\_E1.5  & -16.11  &  21.47   &   6.29 & -16.39 &  22.21 &  7.03 & -17.07 &   24.10 &  8.92 & -17.40 &   24.93 &  9.74 \\
CO3.93\_fm0.3\_E1.5  & -15.81  &  19.45   &   6.67 & -16.28 &  21.50 &  8.73 & -17.07 &   23.52 &  10.74 & -17.42 &   24.86 &  11.05 \\
CO3.93\_fm0.5\_E1.5  & -15.20  &  17.02   &   8.00 & -16.03 &  18.50 &  9.47 & -17.04 &   19.72 &  9.67 & -17.42 &   22.10 &  11.04 \\
CO3.93\_fm0.9\_E1.5  & -15.09  &  10.73   &   6.85 & -15.88 &  12.54 &  7.80 & -16.99 &   17.17 &  11.05 & -17.40 &   18.97 &  11.85 \\
CO3.93\_fm5.0\_E1.5  & -16.07  &  7.25   &   4.01 & -16.27 &  9.03 &  5.11 & -16.99 &   14.88 &  9.81 & -17.31 &   17.04 &  11.08 \\
\hline
CO3.93\_fm0.15\_E1.8  & -16.23  &  20.32   &   5.74 & -16.45 &  21.08 &  7.57 & -17.14 &   23.20 &  9.70 & -17.46 &   23.20 &  9.70 \\
CO3.93\_fm0.3\_E1.8  & -15.91  &  18.47   &   5.93 & -16.33 &  19.06 &  6.52 & -17.14 &   21.52 &  8.98 & -17.46 &   21.52 &  8.98 \\
CO3.93\_fm0.5\_E1.8  & -15.33  &  16.15   &   8.08 & -16.11 &  17.41 &  8.34 & -17.10 &   18.47 &  8.42 & -17.46 &   21.76 &  11.71 \\
CO3.93\_fm0.9\_E1.8  & -15.17  &  9.59   &   5.60 & -15.95 &  13.02 &  8.35 & -17.04 &   15.38 &  9.15 & -17.44 &   17.91 &  10.69 \\
CO3.93\_fm5.0\_E1.8  & -16.08  &  7.16   &   4.12 & -16.29 &  9.04 &  5.29 & -17.03 &   13.72 &  8.90 & -17.34 &   16.13 &  10.55 \\
\hline
CO2.16\_fm0.15\_E1.0  & -17.41  &  10.31   &   4.14 & -16.97 &  11.78 &  5.62 & -17.23 &   13.58 &  7.41 & -17.41 &   13.58 &  7.41 \\
CO2.16\_fm0.3\_E1.0  & -17.38  &  9.51   &   3.88 & -17.03 &  11.14 &  5.50 & -17.34 &   13.11 &  6.47 & -17.50 &   15.15 &  8.50 \\
CO2.16\_fm0.5\_E1.0  & -16.88  &  9.01   &   3.17 & -16.94 &  9.83 &  3.98 & -17.46 &   12.07 &  6.22 & -17.62 &   13.06 &  7.22 \\
CO2.16\_fm0.9\_E1.0  & -16.40  &  7.95   &   3.82 & -16.78 &  9.09 &  4.69 & -17.47 &   11.00 &  6.14 & -17.68 &   11.33 &  6.02 \\
CO2.16\_fm5.0\_E1.0  & -16.32  &  6.82   &   3.75 & -16.72 &  8.06 &  4.63 & -17.44 &   9.74 &  5.61 & -17.69 &   10.73 &  6.31 \\
\hline
\end{tabular}
\tablecomments{
$U_\mathrm{max}$, $B_\mathrm{max}$, $V_\mathrm{max}$, $R_\mathrm{max}$: $U$-, $B$-, $V$-, and $R$-band magnitudes at maximum, 
$t_{U_\mathrm{max}}$, $t_{B_\mathrm{max}}$, $t_{V_\mathrm{max}}$, $t_{R_\mathrm{max}}$: time from the shock breakout
to the $U$-, $B$-, $V$-, and $R$-band maximum, 
$\delta t_{U, -1/2}$, $\delta t_{B, -1/2}$, $\delta t_{V, -1/2}$, $\delta t_{R, -1/2}$: time span from the half maximum to the maximum (i.e., the time needed for a brightness increase by 0.75 mag to maximum)
in $U$-, $B$-, $V$-, and $R$-bands. 
}
\end{center}
}
\end{table*}

\begin{figure*}
\centering
\includegraphics[width=0.999\textwidth]{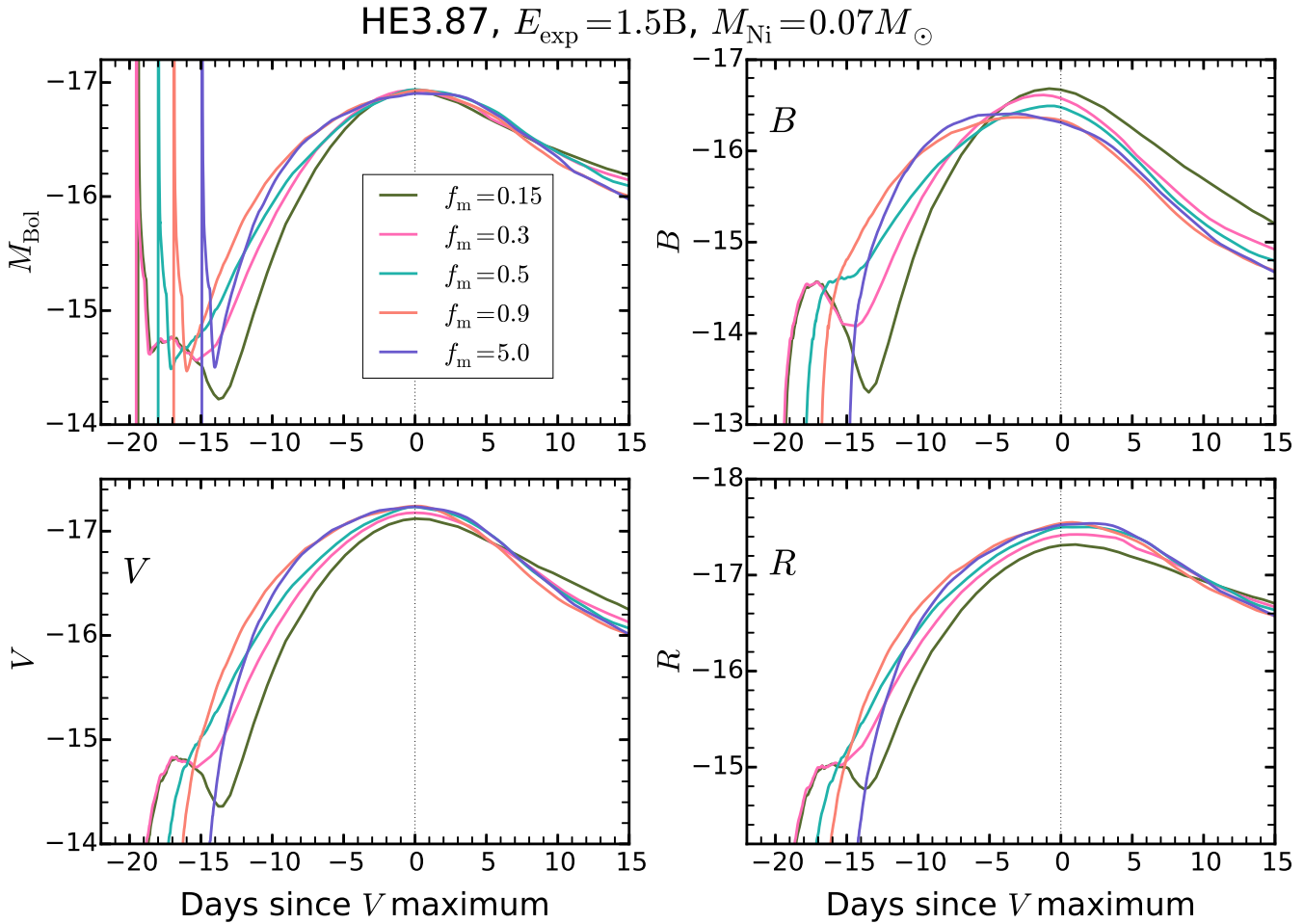}
\caption{Light curves of bolometric, $B$-, $V$- and $R$-band magnitudes of HE3.87 SN models with an explosion energy of $1.5\times10^{51}$~erg 
and a $^{56}$Ni mass of $0.07~M_\odot$. The results with $f_\mathrm{m} =$0.15, 0.3, 0.5, 0.9 and 5.0 are presented using different colors
as indicated by the legends in the upper-left panel.
}  
\label{fig:LCHE}
\end{figure*}

\begin{figure*}
\centering
\includegraphics[width=0.999\textwidth]{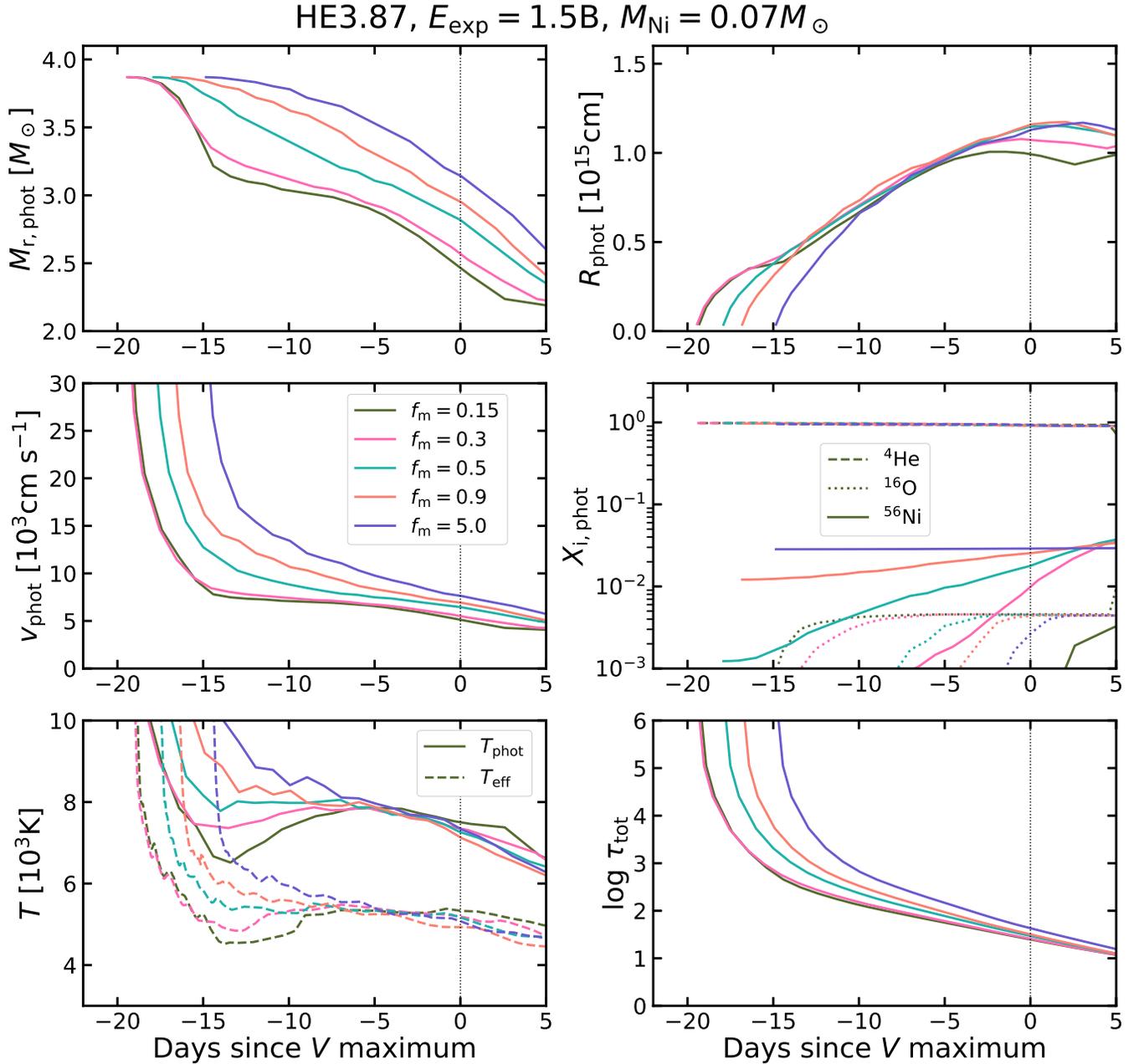}
\caption{Evolution of the photospheric properties of HE3.87 SN models with an explosion energy of $1.5\times10^{51}$~erg 
and  a $^{56}$Ni mass of $0.07~M_\odot$: mass coordinate  (top-left), radius (top-right),   
velocity (middle-left), chemical composition (middle-right), gas temperature (bottom-left) 
at the photosphere. 
The mass coordinate in the top-left panel includes the iron core mass in the progenitor 
(hence the neutron star remnant mass after SN explosion). 
The photosphere is defined 
as the position where the optical depth is 2/3 calculated using the Rosseland-mean opacity. 
In the bottom-left panel, the effective temperatures defined by the Stefan-Boltzmann law 
are also given by the dotted lines for comparison. 
The bottom-right panel gives the total optical depth of the ejecta calculated using the Rosseland-mean opacity. 
The results with $f_\mathrm{m} =$0.15, 0.3, 0.5, 0.9 and 5.0 are presented using different colors
as indicated by the legends in the middle-left panel.
}  
\label{fig:PhotHE}
\end{figure*}

\begin{figure}
\centering
\includegraphics[width=0.999\columnwidth]{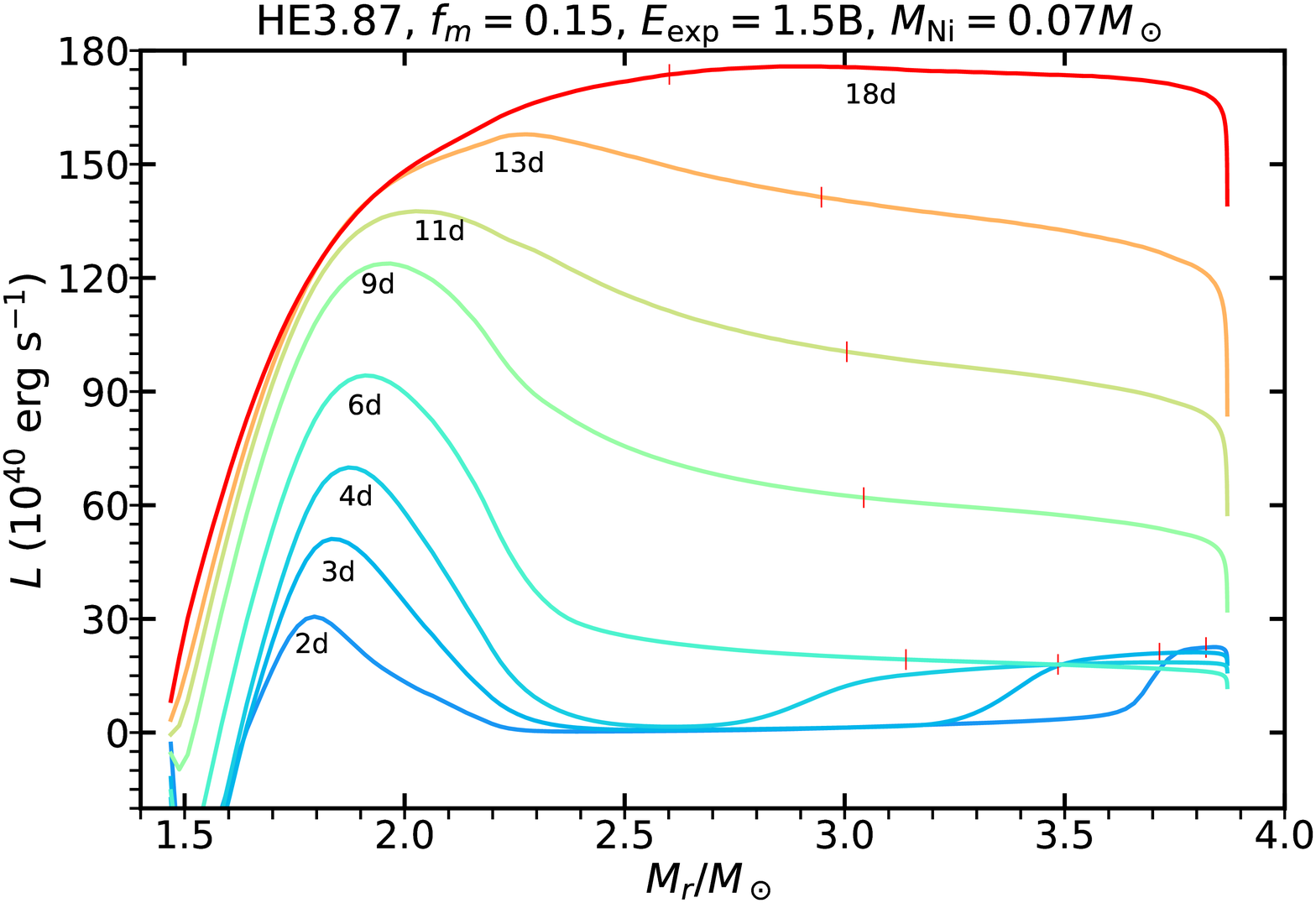}
\includegraphics[width=0.999\columnwidth]{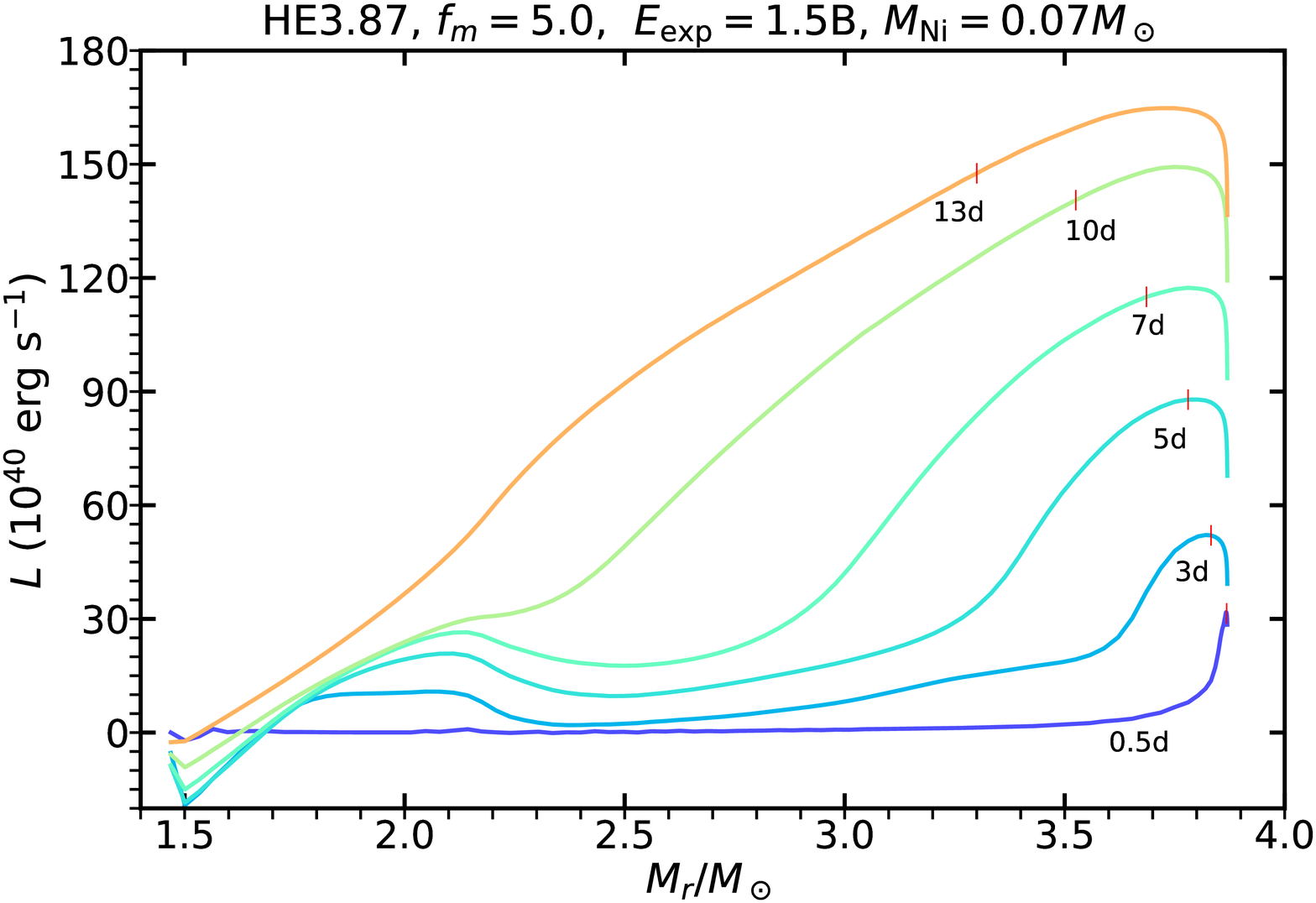}
\caption{Evolution of the luminosity profile in the models HE3.87\_fm0.15\_E1.5 (upper panel)
and HE3.87\_fm5.0\_E1.5 (lower panel). The label in each line denotes days since explosion. 
The vertical red line in each profile gives the position of the photosphere defined with the Rosseland mean opacity.  
}  
\label{fig:lum_wave}
\end{figure}

\begin{figure*}
\centering
\includegraphics[width=0.999\textwidth]{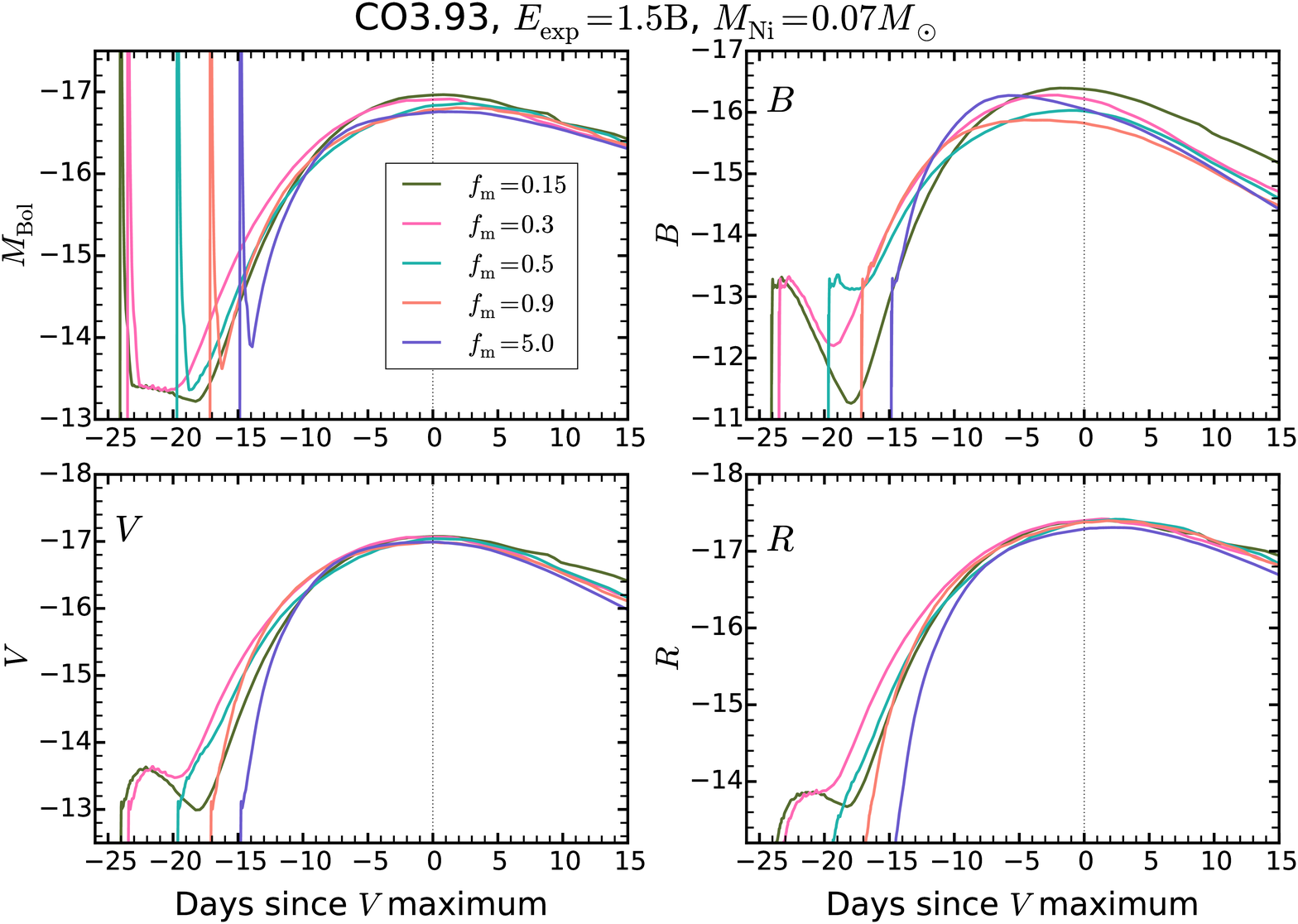}
\caption{Same as in Figure~\ref{fig:LCHE} but for CO3.93 SN models. 
}  
\label{fig:LCCO}
\end{figure*}

\begin{figure*}
\centering
\includegraphics[width=0.999\textwidth]{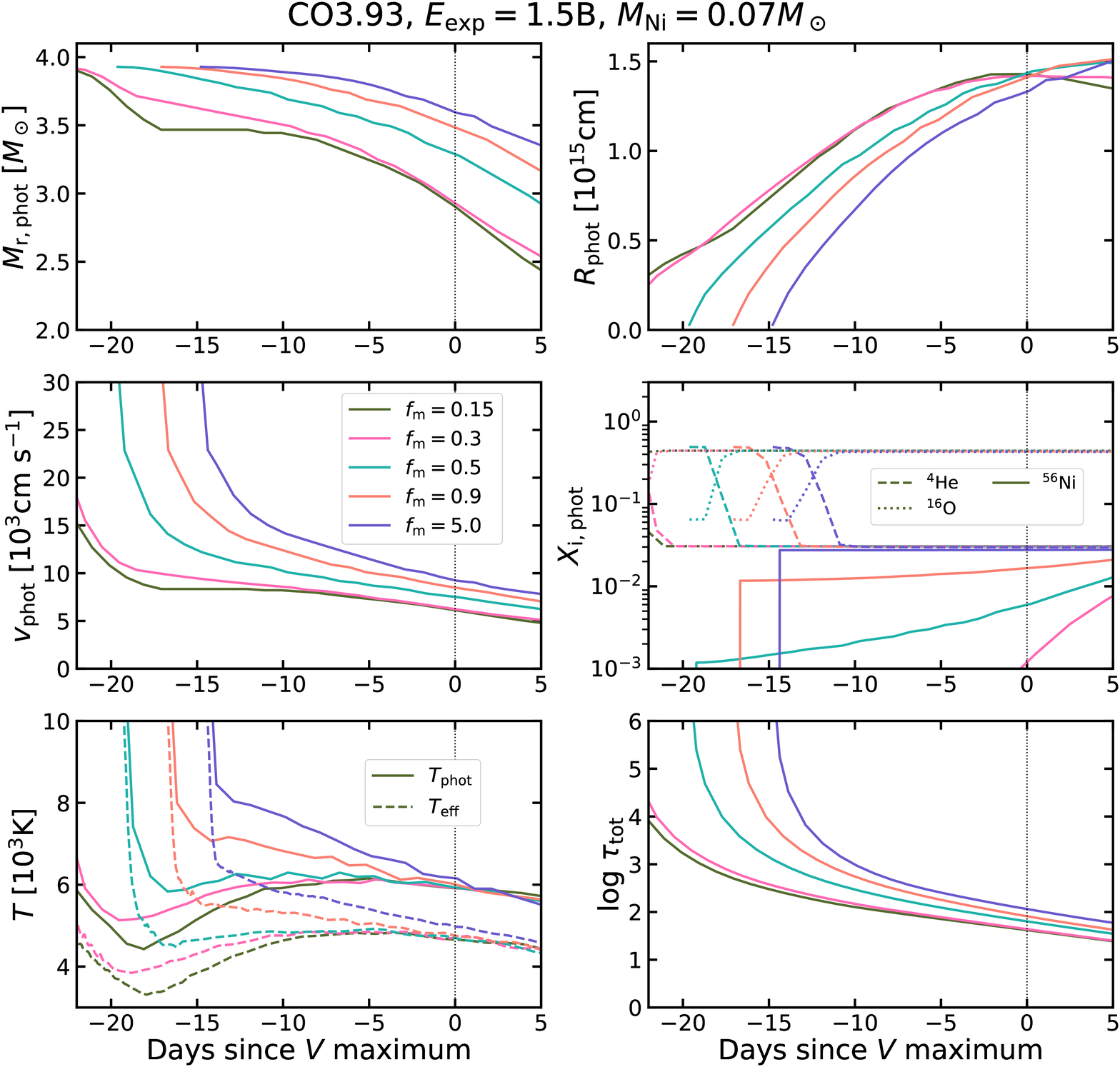}
\caption{Same as in Figure~\ref{fig:PhotHE} but for CO3.93 SN models. 
}  
\label{fig:PhotCO}
\end{figure*}

We use the STELLA code for calculating our SN models.  STELLA is a
one-dimensional Lagrangian multi-group (i.e., energy dependent) radiation
hydrodynamics code \citep{Blinnikov1998, Blinnikov2000, Blinnikov2006} that
implicitly solves the time-dependent radiation transfer equation coupled with
hydrodynamics.  The gamma-ray transfer is solved in a one-group approximation
following \citet{Ambwani1988}. Positron kinetic energy is
absorbed locally (see also \citealt{Blinnikov1998}).  The maximum number of
wave length bins in the code may be up to 1000, but has been fixed to 100 in
our simulations, which covers from $5\times10^4$\AA -- $10^{-3}$\AA{} in
principle.  In most
cases, the shortest wavelength considered in the calculations is about
$0.1$\AA.  This is suitable to properly describe non-equilibrium continuum
radiation.  The ionization is obtained from a solution to the Saha equation
(i.e., it assumes local thermodynamic equilibrium; LTE).  

Our input models are mapped into the STELLA code using a grid of 200 -- 300
mass zones.  The explosion is applied by a thermal bomb at the mass cut with an
explosion energy of 1.0, 1.5 and 1.8$\times10^{51}~\mathrm{erg}$. 

We summarize the properties of the bolometric luminosity of our SN models with
$M_\mathrm{Ni} = 0.07 M_\odot$ in Table~\ref{tab2} and those of $U,B,V,R$-band
magnitudes in Table~\ref{tab3}. The input model, the mixing parameter (i.e.,
$f_\mathrm{m}$) and the explosion energy are indicated in the SN model names of
the tables. For example, HE3.87\_fm0.15\_E1.5 means that the input model is
HE3.87 and that the adopted mixing parameter and explosion energy are
$f_\mathrm{m} = 0.15$ and $E_\mathrm{exp} = 1.5$~B, respectively.  For a given
explosion energy $E_\mathrm{exp}$, the resultant kinetic energy of the SN
($E_\mathrm{K}$) is lower by about $3\times10^{50}~\mathrm{erg}$ for HE3.87
and $4\times10^{50}~\mathrm{erg}$  for CO3.93 and CO2.12
(Table~\ref{tab2}), which correspond to the binding energy above the
mass cut in the progenitor models.  

In the next two sections, we focus our discussion on the SN models with HE3.87
and CO3.93.  The behavior of the SN models with CO2.12 according to different
degrees of $^{56}$Ni mixing is qualitatively similar to that of CO3.93  and we
present the results with CO2.12 only for the comparison with observations, which
is discussed in Section~\ref{sec:comp} below.

\section{Light curves}\label{sec:LC}

From Table~\ref{tab2}, we confirm the following well-known facts about SN
Ib/Ic.  1) The rise time from the shock breakout to the bolometric peak
($t_\mathrm{max}$) and the light curve width (i.e., $\delta t_\mathrm{Bol,+1/2} -
\delta t_\mathrm{Bol,t-1/2}$) become systematically smaller for a higher explosion energy
for a given $f_\mathrm{m}$.  2) The bolometric luminosity at its peak
($L_\mathrm{Bol,max}$) increases with a higher explosion energy for a given
$f_\mathrm{m}$. 

Figures~\ref{fig:LCHE} and~\ref{fig:PhotHE} present the light curves and the
evolution of photospheric properties of HE3.87 SN models with $E_\mathrm{exp} =
1.5\times10^{51}~\mathrm{erg}$.  The photosphere of a a SN Ib/Ic is not well
defined~(see Section 7 of \citealt{Dessart2015}) and in Figure~\ref{fig:PhotHE}
we present properties at the optical depth of 2/3 that is defined with the
Rosseland mean opacity.  In Figure~\ref{fig:LCHE}, the shock breakout
corresponds to the initial spike in the bolometric light curves.  The
subsequent evolution is affected by the $^{56}$Ni distribution.

With a relatively weak mixing of $^{56}$Ni (i.e, $f_\mathrm{m} = 0.15$ and 0.3),
a post-breakout plateau that lasts for several days is observed.  This
short phase is commonly observed in SN Ib/Ic models with weak or no
$^{56}$Ni mixing~\citep{Ensman1988, Shigeyama1990, Dessart2011} and is powered
by the release of shock deposited energy (from both radiation and internal gas energy, 
i.e., ionization energy from helium, carbon and oxygen) as the temperature decreases
continuously.  Readers are
referred to  \citet{Dessart2011} for a detailed discussion on this
post-breakout plateau.  The plateau phase comes to an end when the photosphere
reaches down to the region heated by $^{56}$Ni at about 6~d (the upper panel of
Figure~\ref{fig:lum_wave})  and thereafter the bolometric luminosity starts to
increase towards the main peak. 

In the optical bands, the light curves  with a weak $^{56}$Ni mixing
($f_\mathrm{m} = 0.15$ and 0.3) have double peaks (Figure~\ref{fig:LCHE}).  The
first peak, which is associated with the early post-breakout evolution, is
fainter than the main peak by about 1.6--2.4 mag, depending on the band.  Note
that this magnitude difference between the first and main peaks (in the
optical) would become larger for a higher $^{56}$Ni mass~\citep[which leads to
a brighter main peak; e.g.,][]{Ensman1988}, or for a smaller radius of the
progenitor envelope~\citep[which leads to a fainter first peak;
e.g.,][]{Ensman1988, Dessart2011, Dessart2018}. 

For a stronger  $^{56}$Ni mixing ($f_\mathrm{m} \ge 0.5$), the post-breakout
plateau  practically disappears \citep[see also][]{Dessart2012}.  As shown in
Figure~\ref{fig:PhotHE}, the mass fraction of  $^{56}$Ni at the photosphere
immediately after the shock breakout is already greater than $10^{-3}$ for
$f_\mathrm{m} \ge 0.5$ and $^{56}$Ni heating  plays the dominant role in the
luminosity evolution (see the lower panel of Figure~\ref{fig:lum_wave};
\citealt{Dessart2012, Dessart2015}).  The time span from the shock breakout to
the luminosity peak ($t_\mathrm{Bol, max}$ in Table~\ref{tab2}) becomes shorter
with a stronger $^{56}$Ni mixing, while the light curve width defined by the full
width at half maximum (i.e., FWHM $:= t_\mathrm{Bol, +1/2} - t_\mathrm{Bol,
-1/2}$)  tends to increase instead.  

Figure~\ref{fig:PhotHE} shows that the photosphere (defined here with the
Rosseland mean opacity) is systematically located further out (in radius, velocity,
or Lagrangian mass) for a larger
$f_\mathrm{m}$.  This is mainly because of the change in ionization caused by
the extra heating  in the outermost layers of the ejecta for a higher $^{56}$Ni
abundance.  

The helium-deficient SN models with CO3.93 presented in Figure~\ref{fig:LCCO}
have qualitatively similar properties to those of HE3.87. With $f_\mathrm{m} =
0.15$ and 0.3, the post-breakout plateau is developed as in the case of HE3.87.
Double peak features are also seen in the optical.  The initial radius of
CO3.93 ($R = 0.77~R_\odot$) is much smaller than that of HE3.87 ($R=
6.73~R_\odot$) and the first peaks in the optical are fainter by a factor of a
few than those of HE3.87.  The photosphere in CO3.93 models is systematiclly
located further outward than in HE3.87 models (Figures~\ref{fig:PhotHE} and
\ref{fig:PhotCO}).  The FWHM of the bolometric light curve for a
given set of $f_\mathrm{m}$ and $E_\mathrm{exp}$ is also larger in CO3.93
models (Table~\ref{tab2}), although the CO3.93 and HE3.87 models have similar
kinetic energy and ejecta mass. 

\section{Color evolution}\label{sec:color}

\begin{figure}
\centering
\includegraphics[width=0.99\columnwidth]{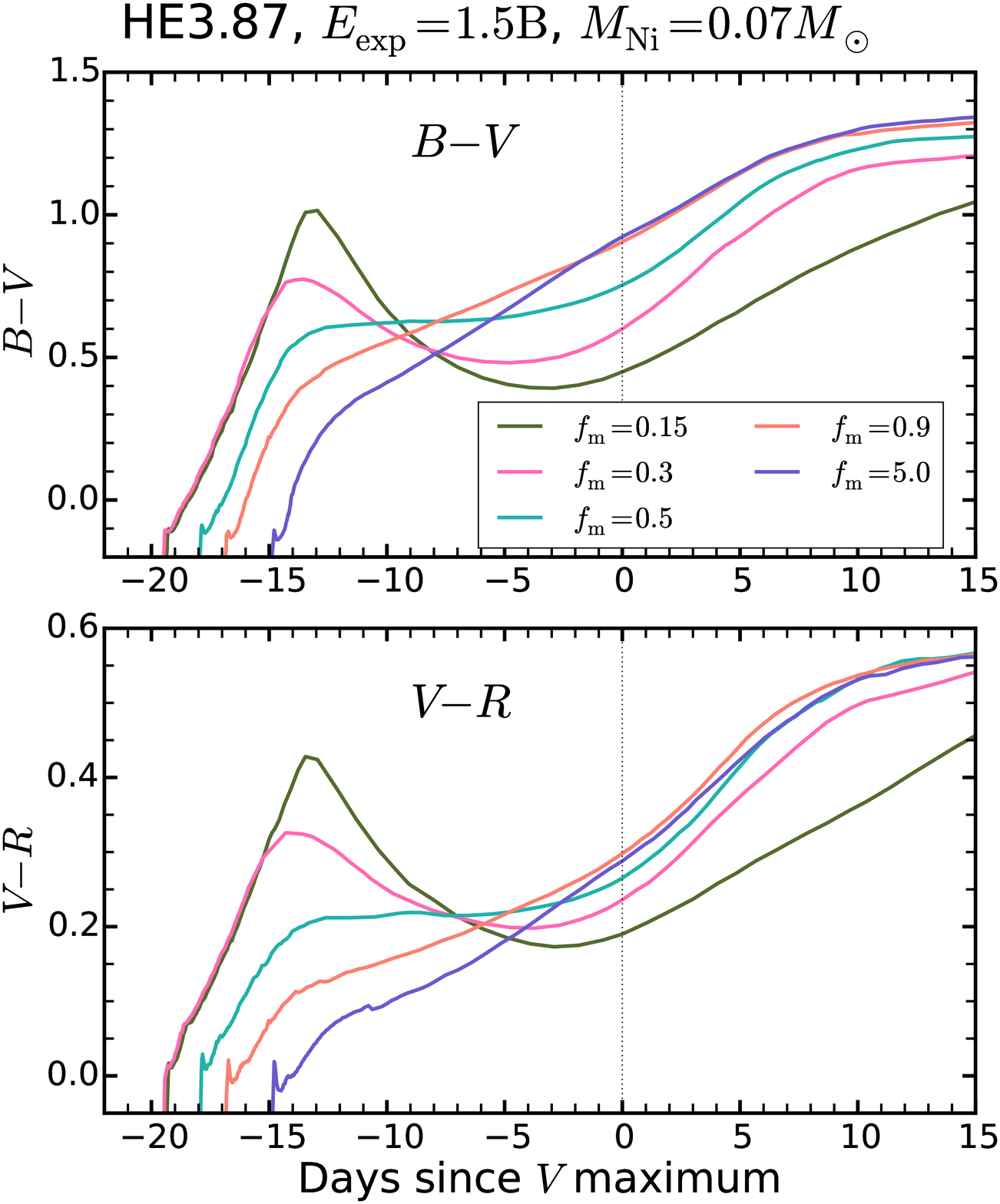}
\caption{Evolution of $B-V$ (upper panel) and $V-R$ (lower panel) colors of HE3.87 SN models with an explosion energy of $1.5\times10^{51}$~erg
and a $^{56}$Ni mass of $0.07~M_\odot$ since 0.05~d after the shock breakout. The results with  $f_\mathrm{m} =$ 0.15, 0.3, 0.5, 0.9 and 5.0 are given by different colors
as indicated by the legeds in the upper panel. 
}  
\label{fig:ColorHE}
\end{figure}

\begin{figure}
\centering
\includegraphics[width=0.99\columnwidth]{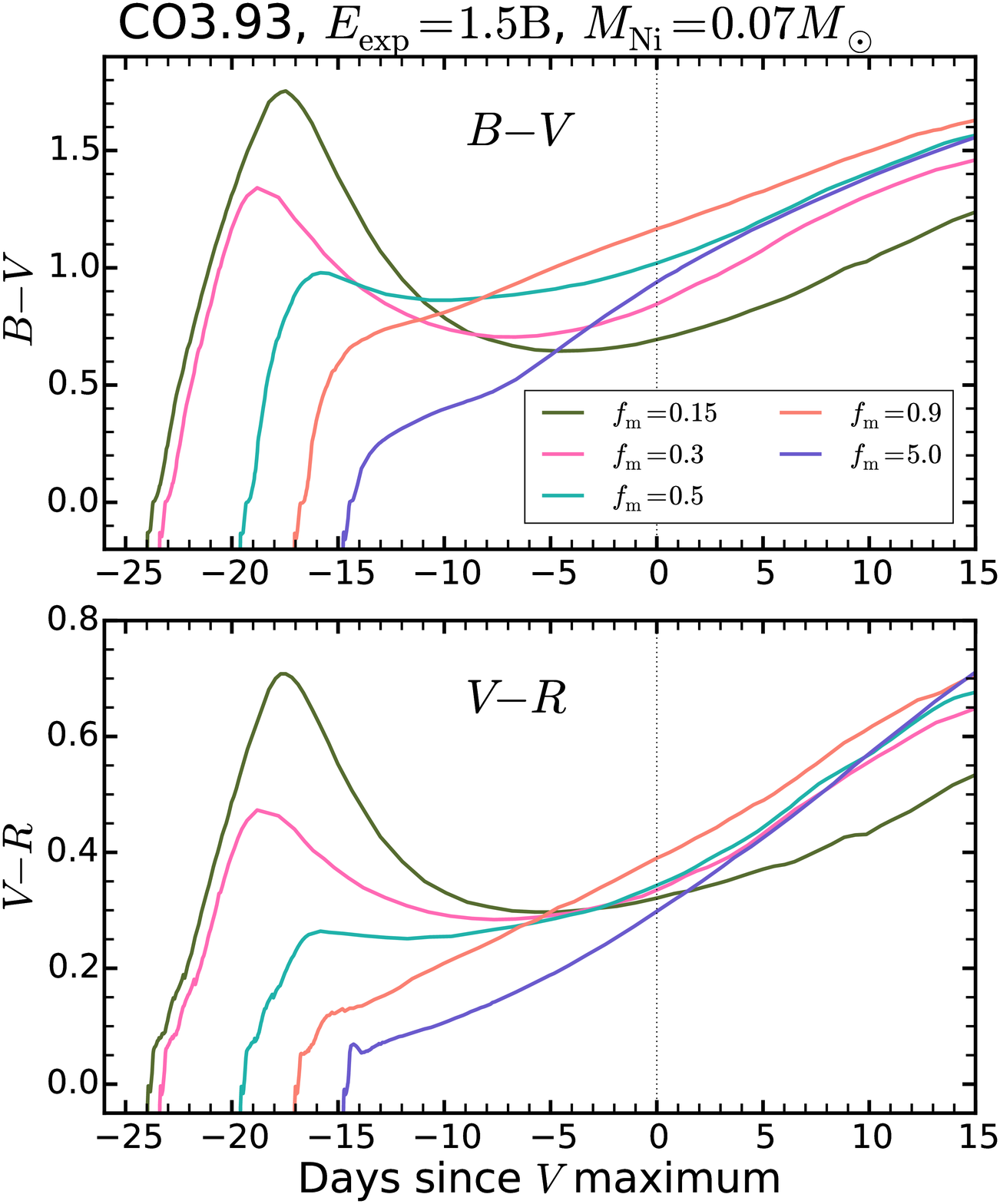}
\caption{ 
Same as in Figure~\ref{fig:ColorHE} but for CO3.93 SN models. 
}  
\label{fig:ColorCO}
\end{figure}

We present the $B-V$ and $V-R$ color evolution of our HE3.87 and CO3.93  SN models with
$E_\mathrm{exp} = 1.5\times10^{51}~\mathrm{erg}$  and $M_\mathrm{Ni} =
0.07~M_\odot$ in Figures~\ref{fig:ColorHE} and~\ref{fig:ColorCO}.   Here the color
evolution is shown only from 0.05~d after the shock breakout because the shock
breakout features, which are practically unobservable
in the optical, make the figure somewhat difficult to read clearly.
The initial evolution is characterized by the rapid increase of $B-V$ and $V-R$
(i.e., the SN reddens) for all the considered $f_\mathrm{m}$ values.
This represents the initial cooling phase due to rapid expansion with no heating source. 

This initial rapid reddening phase is followed by a slope change and the
subsequent evolution is  affected by the $^{56}$Ni distribution.
For a weak $^{56}$Ni mixing ($f_\mathrm{m} = 0.15$ and 0.3),  the color turns
blue-ward at $t \simeq 14$~d and $18$~d before the $V$-band maximum for HE3.87
and CO3.93, respectively, because of the delayed effect of  $^{56}$Ni heating.
This sign change of the slope in the color curves marks the end point of the
post-breakout plateau phase and the beginning of the re-brightening shown in
Figures~\ref{fig:LCHE} and~\ref{fig:LCCO}. This color change in weak mixing
models is also implied by the temperature evolution in Figures~\ref{fig:PhotHE}
and~\ref{fig:PhotCO}, where the photospheric temperature rapidly decreases
initially and starts to increase again.  The color curves reach a local minimum
several days before  the $V$-band maximum and the models continue to redden
thereafter until the nebular phase.  Observations show that the color becomes
blue again $\sim 10$~d after the optical maximum as the SN enters the nebular
phase ~\citep[e.g.,][]{Stritzinger2018}. This phase cannot be properly
described by the STELLA code because the LTE approximation for the gas breaks
down~\citep[cf.][]{Dessart2015, Dessart2016}. 

For a very strong mixing of $^{56}$Ni ($f_\mathrm{m} = 0.9$ and 5.0), the
dominant role of $^{56}$Ni heating makes the initial reddening due to rapid
expansion much weaker and the sign of the color curve slope does not change:
the color continues to redden monotonically. For $f_\mathrm{m} = 0.5$,
which is the intermediate case between weak/extreme mixing, the magnitude
differences after the initial reddening remain nearly constant until a few days
before the $V$-band maximum and increase thereafter.  


\begin{figure}
\centering
\includegraphics[width=0.999\columnwidth]{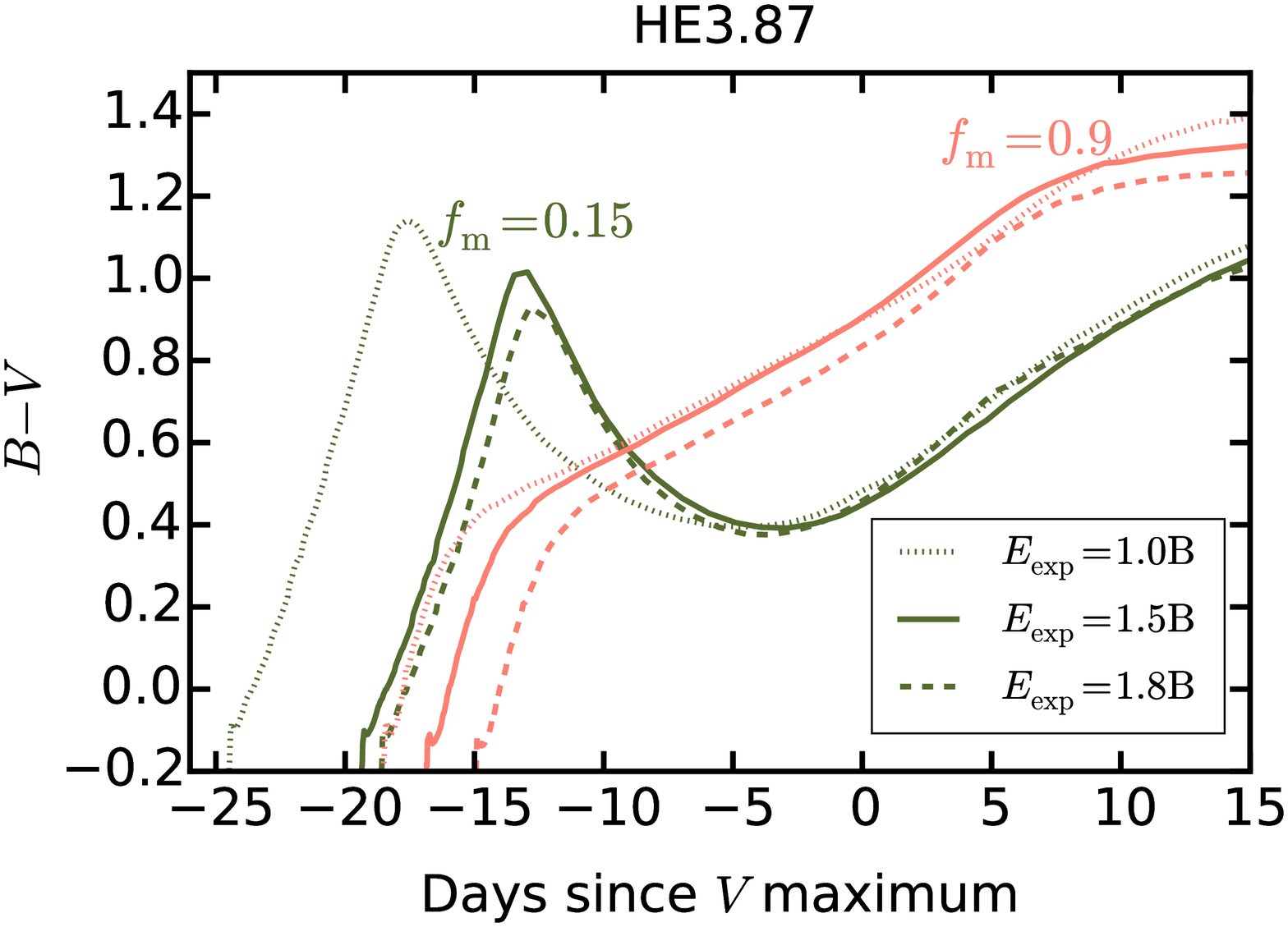}
\includegraphics[width=0.999\columnwidth]{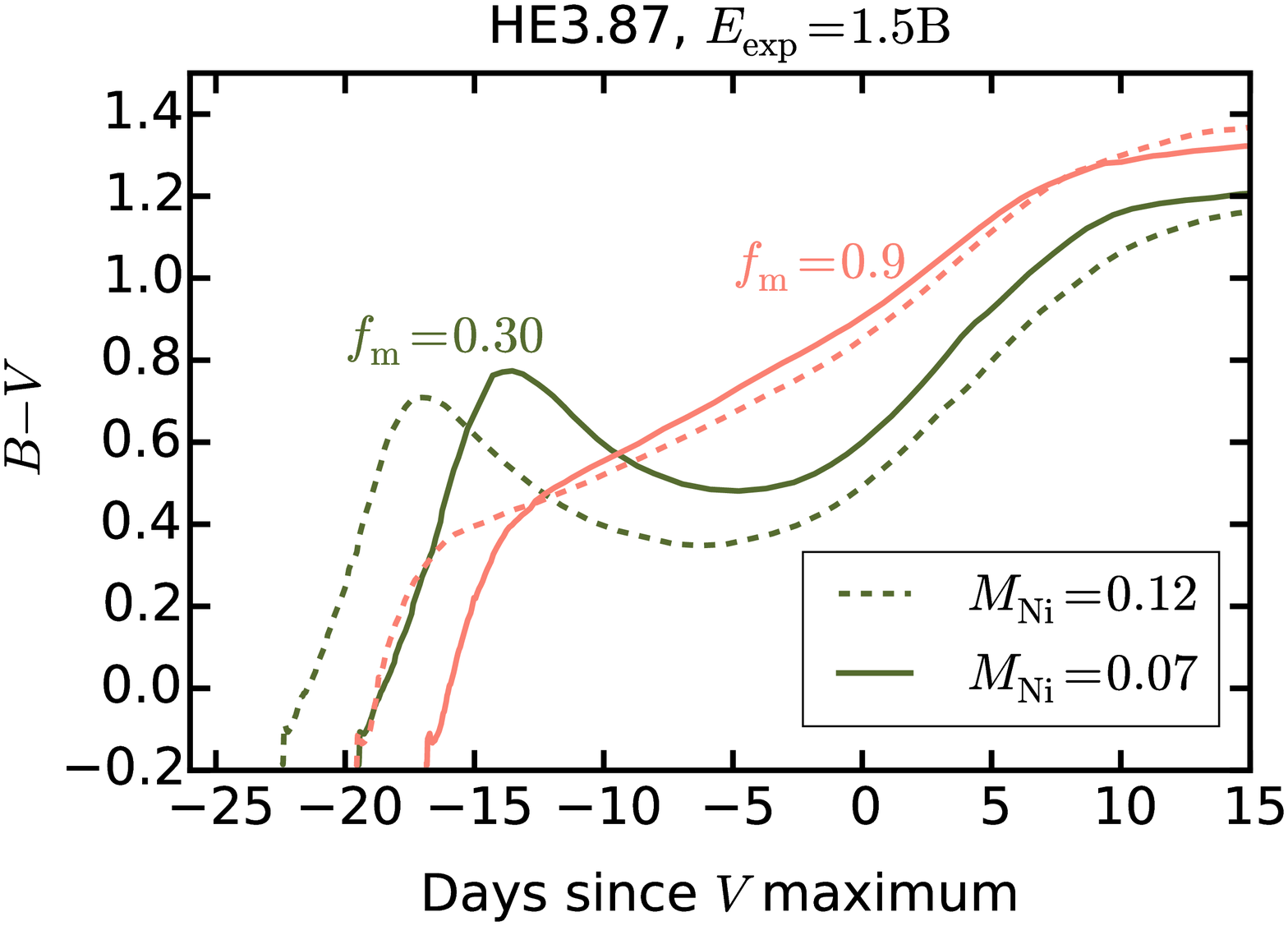}
\caption{Upper panel: the $B-V$ color evolution of HE3.87 SN models with a $^{56}$Ni mass of 0.07~$M_\odot$ and  with $f_\mathrm{m} = 0.15$ (green) and 0.90 (orange)  
for three different explosion energies: $E_\mathrm{exp} =$ 1.0 (dotted), 1.5 (solid) and 1.8 (dashed)$\times10^{51}$~erg.
Lower panel:  the $B-V$ color evolution of HE3.87 SN models with an explosion energy of $1.5\times10^{51}$~erg and with $f_\mathrm{m} = 0.3$ (green) and 0.9 (orange)  
for two different $^{56}$Ni masses: $M_\mathrm{Ni} = 0.07$ (solid line) and $M_\mathrm{Ni} = 0.12$ (dashed line). 
}  
\label{fig:energy}
\end{figure}


In Figure~\ref{fig:energy}, we compare the $B-V$ evolution of HE3.87 models for
different explosion energies.  A lower explosion energy leads to  a longer rise time
to the optical maximum and to a slightly redder color
at the first peak of $B-V$, which marks the end of the initial rapid cooling
phase.  The color at the $V$-band maximum is not significantly affected by the
explosion energy either.  The figure shows that a larger amount of $^{56}$Ni
may make the SN color systematically bluer, which results from more significant
$^{56}$Ni heating, but the overall behavior of the color evolution is not
affected by the total amount of $^{56}$Ni. 

We conclude that the slope change of the color curve is mainly determined by
the $^{56}$Ni distribution.  This means that the color evolution before the
optical maximum would provide a strong observational constraint on 
$^{56}$Ni mixing in SN Ib/Ic as discussed below.

\section{Comparison with observations}\label{sec:comp}

\begin{figure*}
\centering
\includegraphics[width=0.99\columnwidth]{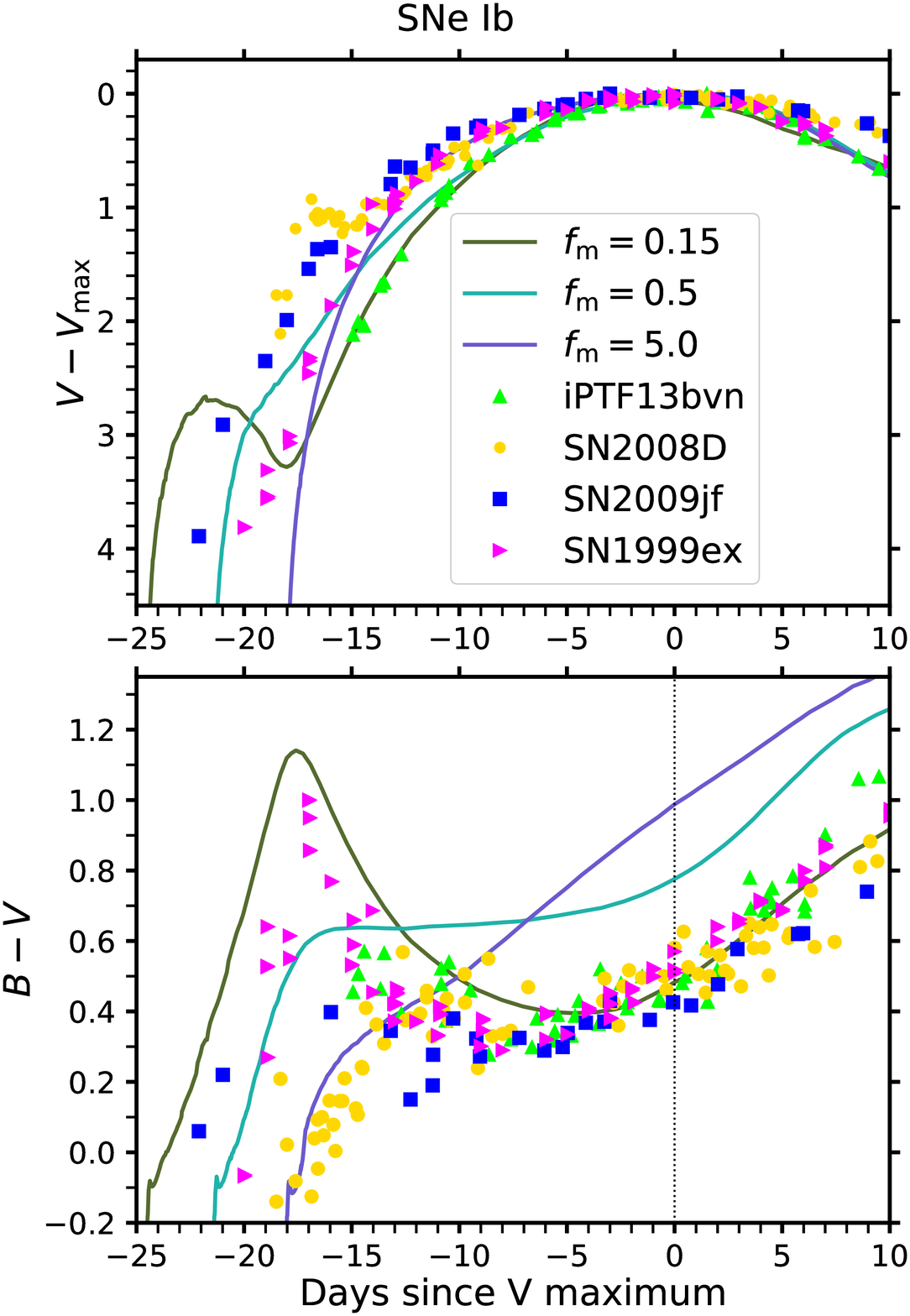}
\includegraphics[width=0.99\columnwidth]{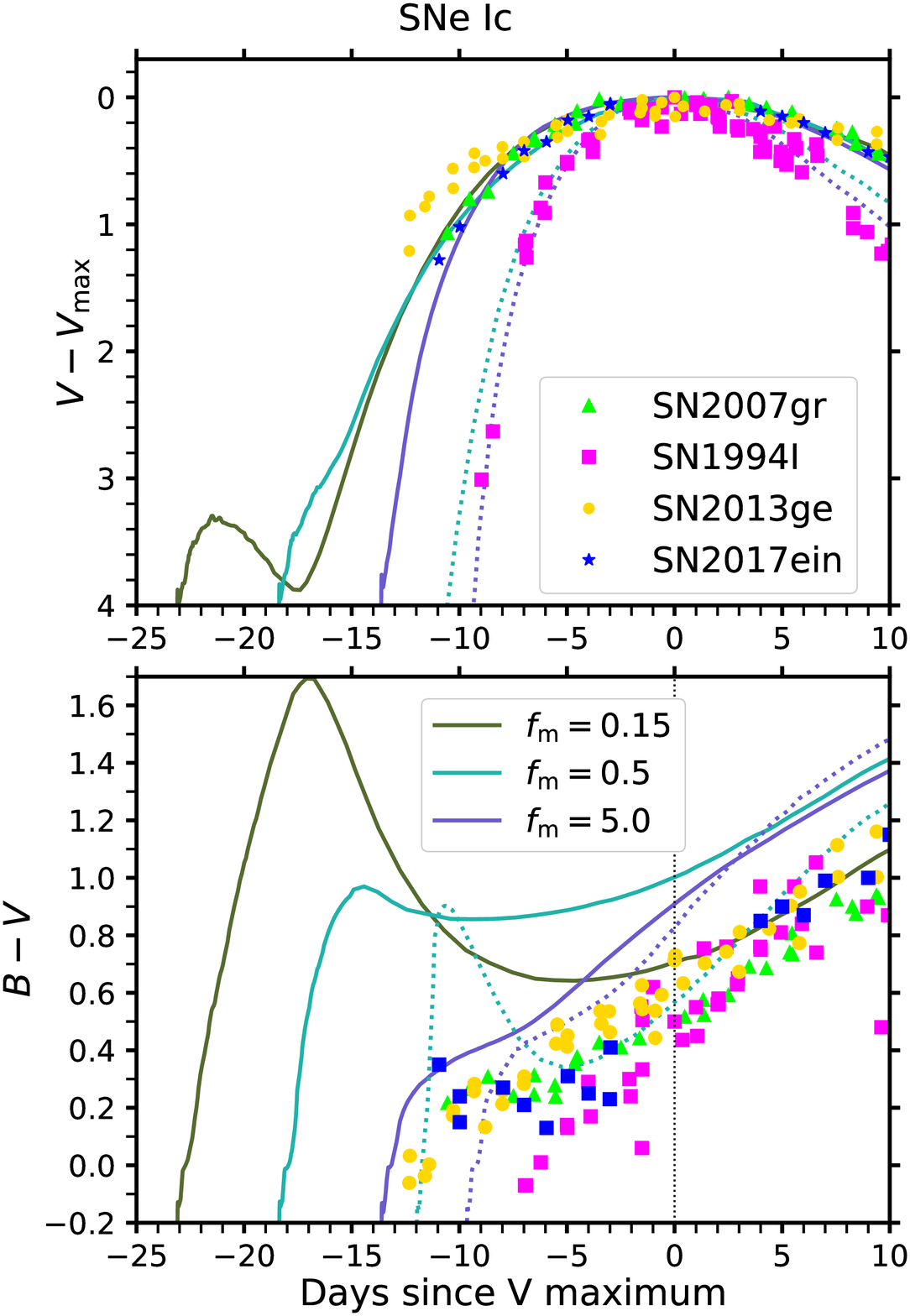}
\caption{Upper left panel: $V$-band light curves with $V_\mathrm{max}$ of iPTF13bvn, SN2008D, SN2009jf, and SN 1999ex compared 
with HE3.87 SN models with $E_\mathrm{exp} = 1.0\times10^{51}$~erg and $M_\mathrm{Ni} = 0.07 M_\odot$. 
Upper right panel: SN 2007gr, SN 2013ge and SN 2017ein compared with CO3.93 SN models with $E_\mathrm{exp} = 1.8\times10^{51}$~erg and $M_\mathrm{Ni} = 0.07 M_\odot$ (solid line), and SN 1994I compared with CO2.16 models with  $E_\mathrm{exp} = 1.0\times10^{51}$~erg and $M_\mathrm{Ni} = 0.07 M_\odot$  (dotted line). 
Lower panel: the corresponding $B-V$ color curves. 
The SN data are taken from the Open Supernova Catalog (\url{https://sne.space}) except  SN 2007gr and SN 2017ein
of which the data are from \citet{Chen2014} and \citet{VanDyk2018}, respectively. Reddening correction ($E(B-V)$) is applied
for each SN as the following. iPTF13bvn: 0.12~\citep{Fremling2016}, SN2008D:  0.6~\citep{Soderberg2008, Modjaz2009}, 
SN2009jf: 0.16~\citep{Valenti2011}, SN 1999ex: 0.28~\citep{Stritzinger2002},  SN 2007gr: 0.085~\citep{Chen2014}, SN 1994I: 0.45~\citep{Richmond1996},  
SN 2013ge: 0.067~\citep{Drout2016}, SN 2017ein: 0.34~\citep{VanDyk2018}.  
}  
\label{fig:compSN}
\end{figure*}

We have shown that the $^{56}$Ni distribution in SN Ib/Ic ejecta impacts their optical color evolution
during the photospheric phase. Here we confront our models with several
observed SNe Ib/Ic. We do not intend to quantitatively infer physical
parameters like the ejecta kinetic energy and mass,  the $^{56}$Ni mass, or the progenitor
radius of each SN.  Instead we focus our discussion on the color
evolution in a qualitative way.  

\subsection{Type Ib supernovae}

On the left side of Figure~\ref{fig:compSN}, we present the $V$-band light
curves and $B-V$ color curves of 4 different SNe Ib (i.e., SN
1999ex,  SN 2008D, SN 2009jf, and iPTF13bvn), compared with our HE3.87 SN
models with $E_\mathrm{exp} = 1.0$~B and $M_\mathrm{Ni} = 0.07 M_\odot$.  These
SNe Ib are very good test cases of our model predictions because they were
observed from about 20 to 15 days before the $V$-band maximum. 

\subsubsection{SN 1999ex}

The most remarkable case is SN 1999ex. As shown in the figure, the initial
rapid reddening phase is unambiguously observed for a period of $t \simeq 20
\cdots 17$~d before the $V$-band maximum. Then the $B-V$ color  reddens until
about $-$10~d before becoming blue again. Here, this behavior is matched by our
weak mixing model (e.g., $f_\mathrm{m} = 0.15$), implying that the $^{56}$Ni
distribution in the SN 1999ex ejecta  is highly concentrated in the center.
The double peak feature predicted in weak mixing models (Figure~\ref{fig:LCHE})
is not found in the $V$-band light curve. However, \citet{Stritzinger2002}
reports a strong signature of the post-breakout plateau in the $U-$band for
this SN. It is likely that the first peak in $V$-band was missed in the SN
survey because it was too faint (i.e., fainter by $ > 4$ mag than at the
optical maximum as implied by Figure~\ref{fig:compSN}).  As discussed in
Section~\ref{sec:LC}, the magnitude difference between the first and main peaks
in the optical can be much larger than what is predicted by the HE3.87 SN
models for a higher $^{56}$Ni mass and/or for a more compact
progenitor~\citep[e.g.,][]{Dessart2018}. A more detailed study should  explain
the lack  of an optical post-breakout plateau  in a quantitative way. 
 
\citet{Hamuy2002} classify SN 1999ex as Type Ib/c instead of Type Ib, arguing
that it is an intermediate case between Ib and Ic. This is because  helium
lines in the optical are relatively weak compared to those of ordinary SNe Ib.
The weak $^{56}$Ni mixing evidenced by the color evolution of SN 1999ex implies
that the weakness of \ion{He}{1} lines is not necessarily due to a relatively
small amount of helium. Note also that \ion{He}{1} lines of this SN gradually
become stronger as the light curve approaches its main peak as in the case of
ordinary SNe Ib~\citep{Hamuy2002}.  This is consistent with the case of weak
$^{56}$Ni mixing in a helium-rich ejecta  where thermal processes form weak
\ion{He}{1} lines initially and in later stages non-thermal processes due to
radioactive decay of $^{56}$Ni gradually become more important to make stronger
\ion{He}{1} lines~\citep{Dessart2012}. 

We conclude that the progenitor of SN 1999ex might have retained a fairly large
amount of helium and that the relative weakness of \ion{He}{1} lines are not
necessarily due to a small content of helium.   

\subsubsection{iPTF13bvn}

The color evolution in Figure~\ref{fig:compSN} indicates that the initial rapid
reddening phase is missing for iPTF13bvn.  However, we observe that the color
reddens from 18~d until about $13$~d before the $V$-band maximum and becomes
bluer thereafter until $7$~d before the $V$-band maximum. This provides
evidence for a relatively weak $^{56}$Ni mixing.  The difference in the
strength of \ion{He}{1} lines between iPTF13bvn and SN 1999ex would have
resulted from somewhat different degrees of $^{56}$Ni
mixing~\citep[cf.][]{Dessart2012}. For example, both HE3.87 SN models with
$f_\mathrm{m} = 0.15$ and $f_\mathrm{m} = 0.3$ have a similar qualitative slope
change in the color evolution (Figure~\ref{fig:ColorHE}),  but the influence of
$^{56}$Ni radioactive decay at the photosphere for the formation of \ion{He}{1}
lines  would become more important with $f_\mathrm{m} = 0.3$ as implied by the
evolution of the chemical composition at the photosphere presented in
Figure~\ref{fig:PhotHE}.

\subsubsection{SN 2009jf}

Early time data is available for SN 2009jf (i.e., $V-V_\mathrm{max}  \simeq
4$~mag; Figure~\ref{fig:compSN}).  The initial rise in the light curve from
$V-V_\mathrm{max} \simeq 4$ mag to $V-V_\mathrm{max} \simeq 3$~mag  is very
rapid but a small decrease in the slope is observed from $V-V_\mathrm{max}
\simeq 3$ to $V-V_\mathrm{max} \simeq 2.4$~mag.  This behavior is qualitatively
similar to the HE3.87 SN model prediction with $f_\mathrm{m} = 0.5$.  The color
evolution  is also qualitatively consistent with the prediction with
$f_\mathrm{m} = 0.5$, although the model prediction is much redder than the
observation.  Unfortunately, however,  the $B-V$ data is missing for $t \simeq
= 19 \cdots 17$~d before the $V$-band maximum, and we do not know how strong
the initial reddening was.  This makes it difficult to precisely determine
which model is most consistent with the color evolution of this SN. 

\subsubsection{SN 2008D}

SN 2008D is a SN Ib having very luminous post-breakout
emission~\citep{Soderberg2008, Malesani2009, Modjaz2009}.  \citet{Bersten2013}
invoked jet-induced $^{56}$Ni mixing into high-velocity outermost layers of the
ejecta to explain the early time evolution of this SN.  More recently,
\citet{Dessart2018} argued that 
the featureless spectra having weak \ion{He}{1} lines during early times can
be better explained by a helium-giant progenitor having an extended and tenuous
envelope (i.e., $R \sim 200 R_\odot$).  The post-breakout plateau phase is
rather short (i.e., about 3 days) and is followed by an increase in
luminosity towards the main peak.  

In the lower panel of Figure~\ref{fig:compSN} we see that SN 2008D does not
become as red as SN 1999ex at the end of the initial reddening phase 
(i.e., $B-V = 0.5 - 0.6$~mag at $t \simeq -13$~d for SN 2008D
compared to $B-V \simeq 0.1$~mag at $t \simeq -17$~d for SN 1999ex). 
In the context of our study on $^{56}$Ni mixing, this would imply a stronger mixing 
in SN 2008D than in SN
1999ex, although not as extreme as assumed by \citet{Bersten2013}.
This property of relatively strong ejecta mixing may be compatible with the giant progenitor 
model proposed by \citet{Dessart2018}, which is crucial to explain the large early-time 
brightness. Indeed, in the helium giant scenario, chemical mixing induced by the Rayleigh-Taylor
instability would be more efficient~\citep{Shigeyama1990, Hachisu1991}.  
Subsequently, the color behavior is quite similar to the other events and very
strong mixing (i.e., $f_\mathrm{m} = 0.9 - 5.0$) is ruled out.  

\subsection{Type Ic supernovae}

The properties of ordinary SNe Ic, as we understand them today, 
are similar to those of SNe Ib in terms of
kinetic energy, ejecta mass, and $^{56}$Ni mass in general. The observed
sample is heterogeneous, with a greater diversity for SNe Ic relative to SNe Ib. 
For example, SN 1994I,
which is the prototype of SNe Ic, has an unusually short rise time.  
Furthermore, all  SNe associated with long gamma-ray bursts (GRB) and all
hydrogen-deficient hyper-energetic/superluminous SNe  have been found to belong
to Type Ic~\citep{Woosley2006, GalYam2012, Branch2017}. So far, there is no SN Ib
associated with a long GRB or hyper-energetic/superluminous SN,
with the possible exceptions of SN 2016coi, which exhibits
unusually broad \ion{He}{1} lines~\citep{Yamanaka2017}, and of SN 2005bf, whose 
high luminosity may be powered by a magnetar~\citep[e.g.,][]{Maeda2007}.
The origin of the SN Ic
diversity is currently poorly known. Rapid core rotation in the SN progenitor may be a
necessary condition for making a central engine (either collapsar or
magnetar) susceptible to power a hyper-energetic/superluminous SN~\citep{Woosley1993,
Wheeler2000}.  Given that our 1-D STELLA models cannot be directly compared to
these peculiar SNe Ic, whose ejecta are probably highly asymmetric
 ~\citep[e.g.,][]{Maeda2003, Dessart2017b},
our discussion is limited to SNe Ic that appear to be ordinary in terms of
kinetic energy ($E_\mathrm{K} = (1-2)\times10^{51}$~erg) and 
$^{56}$Ni mass ($M_\mathrm{Ni} \approx 0.1 M_\odot$).

On the right side of Figure~\ref{fig:compSN} our CO3.93 SN models with
$E_\mathrm{exp} = 1.8$~B and $M_\mathrm{Ni} = 0.07 M_\odot$ are compared with
SN 1994I, SN 2007gr, SN 2013ge and SN 2017ein. The early-time features  of
these SNe are relatively well studied compared to those of other SNe Ic. The
first data for SN 1994I is very deep ($V-V_\mathrm{max} \simeq 3.0$) but for
the other three SNe Ic, we have only $V-V_\mathrm{max}\approx 1.0$ at the first
data point,  which is much smaller than in the case of SNe Ib (i.e.,
$V-V_\mathrm{max}\gtrsim 2.0$ at the first data point).  Therefore, our argument 
presented below for these three SNe Ic is weaker than in the case of SNe Ib. 

There is no clear evidence for \ion{He}{1} lines 
in the optical spectra of SN 1994I,  SN 2007gr and SN 2017ein, which have been 
classified as Type Ic.
\citet{Drout2016} report weak \ion{He}{1} lines in the early
optical spectra of SN 2013ge and classify this SN as SN Ib/c.  However, unlike
SN 1999ex, these \ion{He}{1} signatures soon disappear and SN 2013ge would have
been classified as SN Ic without early-time spectra.  This suggests
that signatures of helium could have been observed in many other SNe Ic if
data at earlier times had been available.  The color evolution of SN 2013ge,
which shows a monotonic increase in $B-V$, is qualitatively different
from the SNe Ib discussed above but similar to SN 1994I.   

The ejecta masses of SN 2007gr, SN2013ge and SN 2017ein are inferred to be
about $2-3~M_\odot$~\citep{Valenti2008, Mazzali2010, Drout2016, VanDyk2018},
which is comparable to that of the CO3.93 model. The caveat is that these
estimates based on the light curves are subject to  
uncertainties~\citep[e.g.,][]{Dessart2016}.  Non-detection with a very
deep limit is reported 5 and 7~d before the discovery of SN 2007gr and SN
2013ge, respectively~\citep{Valenti2008, Drout2016}.  SN 1994I has a very low
ejecta mass (i.e., $M_\mathrm{ej} \approx 0.6 - 1.0 M_\odot$; e.g.,
\citealt{Nomoto1994, Iwamoto1994, Sauer2006}) that leads to a much narrower
light curve than those of SN 2007gr and SN 2013ge.  The first detection of SN
1994I is likely to be within a few days from explosion~\citep{Richmond1996}.

As a caveat, note that the observed $B-V$ color of our SN Ic sample is
systematically bluer than the model predictions.  This difference might be
attributed to different $^{56}$Ni/ejecta masses, uncertainties in the reddening
correction, non-LTE effects that are not included in our STELLA models,
or our mixing procedure (which affects only the $^{56}$Ni distribution).
Observations indicate significant asphericity in ordinary SNe Ib/Ic
\citep[e.g.,][]{Maund2007, Tanaka2008, Tanaka2012,  Reilly2016, Tanaka2017} and
multi-dimensional effects might also play an important role in the color
evolution.

\subsubsection{SN 1994I and SN 2013ge}

As shown in Figure~\ref{fig:compSN} (the bottom right panel), the $B-V$
value of SN 1994I and SN 2013ge  monotonically increases from
the first detection.  Given that the time span between the non-detection and
the discovery of these SNe is fairly short ($< 1-2$~d for SN1994I and $< 5$~d
for SN 2013ge), it is unlikely that  the signature of delayed $^{56}$Ni
heating (i.e., decrease of $B-V$  after the initial reddening phase) could have
been present before the discovery.  Therefore, the color evolution 
implies that $^{56}$Ni is almost fully mixed in the ejecta of these SNe.  This
contrasts with the case of SNe Ib, for which the color behavior points 
to a relatively weak $^{56}$Ni mixing.  

As shown by \citet{Hachinger2012} and \citet{Dessart2012}, not much helium
could be hidden in SNe Ic if $^{56}$Ni were strongly mixed in the ejecta\footnote{The
value $m_\mathrm{He} < \sim 0.2 M_\odot$ has been proposed for SN 1994I, for which a very
low ejecta mass has been inferred, but the situation
is probably more complicated in general. The presence of \ion{He}{1} lines may also 
depend on the CO core mass, the local He mass fraction, the level of chemical 
segregation and such.}. Therefore, our analysis leads to the
conclusion that the progenitors of SN 1994I and SN 2013ge did not
have a helium-rich envelope, being an almost naked carbon-oxygen core. 

\subsubsection{SN 2007gr and SN 2017ein}

For SN 2007gr, the $B-V$ color curve is somewhat flat for $t = - 10 \cdots
-5$~d since the $V$-band maximum, and it seems to be compatible with the
case of $f = 0.5$.  Given the lack of earlier data for this SN, however, we
cannot well constrain the degree of $^{56}$Ni mixing based on the color evolution. 

SN 2017ein has a weak signature of the effect of delayed $^{56}$Ni heating on the
$B-V$ color curve (i.e., the slight decrease of $B-V$ during $t = -12 \cdots -10$~d
since the $V$-band maximum).  This indicates that $^{56}$Ni mixing in this SN was
 weaker than in SN 1994I and SN 2013ge.  Although
the lack of \ion{He}{1} lines around the $V$-band maximum might be due to
weak  $^{56}$Ni mixing, the constraint set on He is weak given the lack of spectra 
within $\sim$\,5 d of explosion. At such times,
\ion{He}{1} lines may be  produced even without non-thermal effects if the 
the progenitor had a massive helium-rich envelope with a very high helium mass fraction
(i.e., $X_\mathrm{He} \gtrsim 0.9$; \citealt{Dessart2011}).

\section{Implications for SN Ib/Ic progenitors and nickel mixing}\label{sec:implication}

\begin{figure*}
\centering
\includegraphics[width=0.7\textwidth, angle=90]{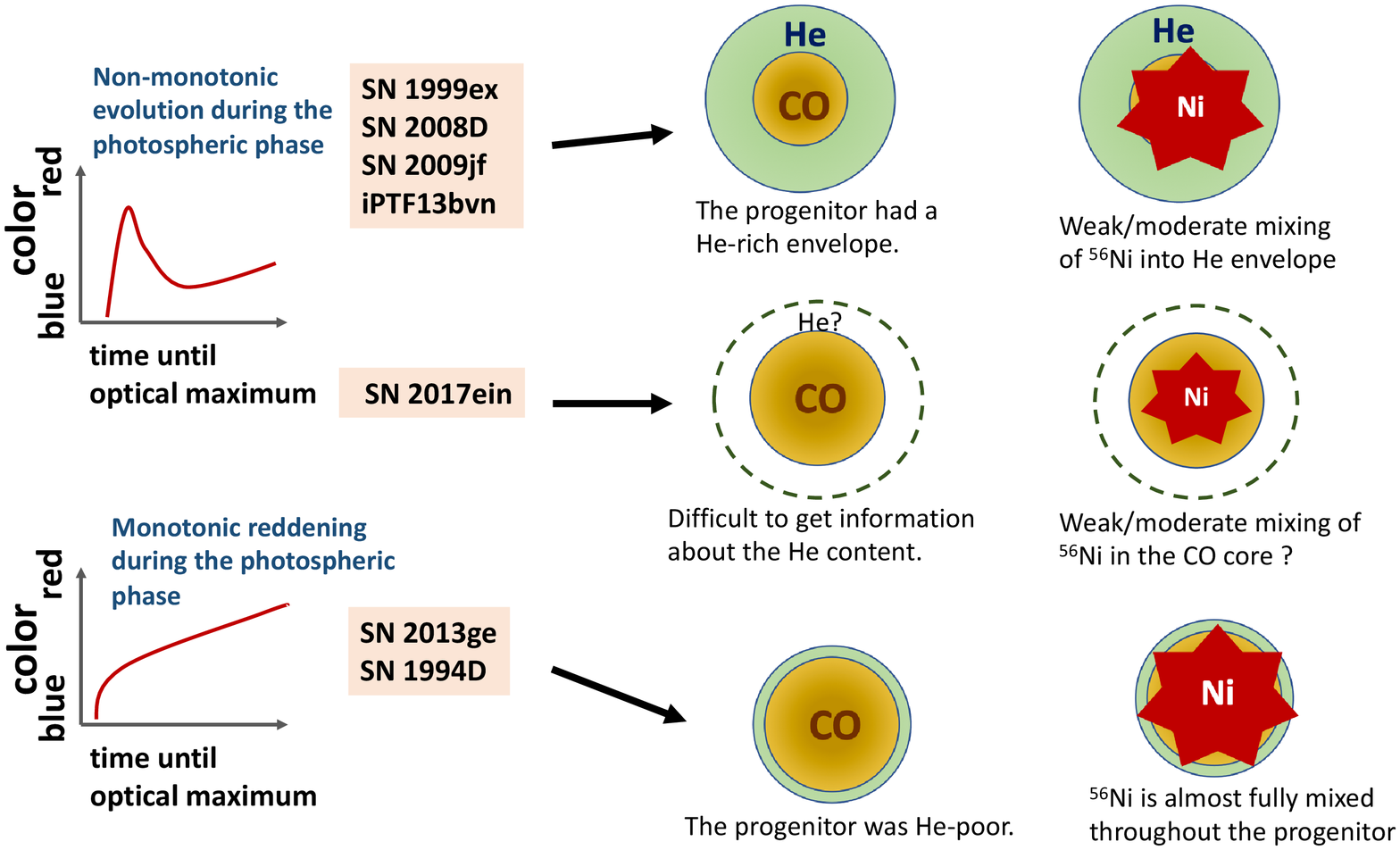}
\caption{Schematic summary of our discussion in Section~\ref{sec:comp}.  
}  
\label{fig:summary}. 
\end{figure*}
 
We schematically summarize our discussion of the previous section in
Figure~\ref{fig:summary}. We tentatively conclude that the color evolution of
SNe Ib differs from that of SNe Ic because of different $^{56}$Ni distributions
in the ejecta.  For SNe Ib, strong mixing of $^{56}$Ni into the
helium-rich envelope is ruled out and only relatively weak/moderate mixing can
be compatible with observations, while very weak mixing is also ruled out as
otherwise no \ion{He}{1} lines would form at the optical maximum.  For some SNe
Ic (i.e., SN 1994I and SN 2013ge), $^{56}$Ni seems to be strongly mixed
throughout the SN ejecta, implying that the progenitors of these SNe Ic are
intrinsically helium poor as otherwise helium could not be hidden in the
spectra. 

From this finding, we suggest the following propositions regarding SN Ib/Ic progenitors and $^{56}$Ni mixing
in SNe, which need to be tested by future studies:
\begin{enumerate} 
\item Progenitors of SNe Ib and SNe Ic do not form a continuous sequence in terms of helium content in general.
They differ  for most cases
meaning that SNe Ib progenitors have a helium-rich envelope with a high mass fraction of helium ($\gtrsim 0.9$) 
and SNe Ic progenitors are almost naked CO cores that
only have a small amount of helium left in the outermost layers with a low mass fraction of helium. 
\item Mixing of $^{56}$Ni into the CO core during the explosion is very efficient irrespective of SN type.
Hence, type Ic SNe, whose progenitors are naked CO cores,  exhibit a monotonic (optical) color evolution
until the nebular phase, which is a signature of strong $^{56}$Ni mixing.
\item In progenitors having a helium-rich envelope, mixing of $^{56}$Ni from 
the CO core into the helium envelope induced by hydrodynamic instabilities  is probably weaker,
depending on the progenitor structure.
\end{enumerate}

The first proposition needs to be confronted to the predictions of stellar
evolution theory.  Type Ib/Ic SN progenitor models by \citet{Yoon2010} predict
that the total helium mass in SN Ib/Ic progenitors decreases in a continuous
way as a function of the initial mass.  This is  a consequence of the adopted
mass-loss rate prescription for Wolf-Rayet (WR) stars given by
\citet{Hamann1995} that depends only on the luminosity and the surface hydrogen
abundance.  Recently \citet{Yoon2017b} revisited the mass-loss rate
prescription of Wolf-Rayet (WR) stars.  He found that the helium mass
distribution in SN Ib/Ic progenitors would be bimodal, i.e., either larger than
about $1.0~M_\odot$ or less than about $0.2~M_\odot$ depending on the initial
mass of the progenitors, if the mass-loss enhancement during the WC stage of WR
stars compared to the case of WN stage were properly taken into account.
Therefore, the first proposition is supported by the conclusion of
\citet{Yoon2017b}\footnote{SNe Ic like SN 1994I  have a very small ejecta mass
($M_\mathrm{ej} \lesssim 1.0~M_\odot$). Mass transfer from naked helium stars
during core carbon burning in close binary systems (so-called Case BB) would
play an important role for their progenitors~\citep{Nomoto1994, Wellstein1999,
Yoon2010, Tauris2015}.}.  This is also consistent with the recent observations
that indicate systematic differences between SNe Ib and Ic in spectral
properties~\citep{Liu2016, Prentice2017}.  For example, it is found that SNe Ic
have a stronger absorption line \ion{O}{1}~$\lambda7774$ and higher velocities
of \ion{Fe}{2}~$\lambda5169$ and \ion{O}{1}~$\lambda7774$ than SNe
Ib~\citep{Liu2016}.

The second proposition seems to be well supported by various multi-dimensional
simulations where efficient $^{56}$Ni mixing into the CO core is often
observed~\citep[e.g.,][]{Kifonidis2003, Kifonidis2006, Hammer2010, Ono2013,
Mao2015, Wongwathanarat2017, Mueller2018}.  Recently \citet{Taddia2018} found
that SNe Ic light curves are particularly well fitted by  models with a
significant $^{56}$Ni mixing, which also supports the second proposition. 

Mixing of $^{56}$Ni into the helium envelope would depend on the progenitor
structure as discussed by \citet{Shigeyama1990} and \citet{Hachisu1991}.  Using
2D numerical simulations, \citet{Hachisu1991} showed that  $^{56}$Ni mixing
into the helium envelope due to the R-T instability in a SN progenitor having a
helium-rich envelope   can become more efficient for lower mass progenitors
because of the higher density contrast between the CO core and the helium
envelope.  Almost complete mixing of $^{56}$Ni into the helium envelope as in
the cases of $f_\mathrm{m} = 0.9$ and 5.0 of our models
(Figure~\ref{fig:initial}) is not found in their simulations.  In their 3.3
$M_\odot$ helium star model where mixing occurs most efficiently in their
calculations, $^{56}$Ni is mixed only up to a middle point of the helium
envelope, which is comparable to the cases of $f_\mathrm{m}=0.3 - 0.5$ in our
calculations. Such moderate mixing is suitable for explaining the color
evolution of SN 2008D and SN2009jf. For 6 $M_\odot$ helium star model, on the
other hand, the R-T instability is shown to be very week and mixing occurs only
in a limited region around the interface between the CO core and the helium
envelope, which may well explain the properties of SN 1999ex.  Therefore, the
third proposition above is in good agreement with the result of
\citet{Hachisu1991}.  However, Hachishu et al. did not consider the effects of
neutrino-driven convection and $^{56}$Ni fingers that play an important role
for chemical mixing in SN ejecta \citep[e.g.,][]{Kifonidis2003}.  It is also
likely that the degree of $^{56}$Ni mixing depends on the ratio of the
$^{56}$Ni mass to the SN ejecta mass (i.e., weaker mixing for a lower $^{56}$Ni
to ejecta mass ratio).  Further systematic investigations of chemical mixing in
SNe Ib/Ic with more realistic multi-dimensional simulations are needed.

\section{Conclusions}\label{sec:conclusion}

We have discussed the effects of the $^{56}$Ni distribution in SN ejecta on the
early-time light curve and color evolution of SNe Ib/Ic.  The results presented
in Section~\ref{sec:LC} imply that the presence or absence of an early plateau
phase can help constrain the degree of $^{56}$Ni mixing.  However, such an
early plateau, if present, would be short-lived and much fainter than the main
peak and would thus be easily missed.  A complementary and more suitable
approach is to use the early-time color evolution.  We have shown that it can
be a sensitive diagnostic of the $^{56}$Ni distribution
(Section~\ref{sec:color}).  With a weak $^{56}$Ni mixing, the early-time
optical color initially reddens for about a week (radiation and expansion
cooling is not compensated by $^{56}$Ni heating in photospheric layers),
subsequently evolves to the blue up to a week before maximum (delayed heating
from $^{56}$Ni), followed by a continuous reddening until the nebular phase.
With a strong $^{56}$Ni mixing, the effect of $^{56}$Ni heating is more
continuous and progressive, so that the color is initially bluer than in weakly
mixed cases and the color reddens slowly and monotonically during the
photospheric phase.  

We have shown that the relatively weak/moderate $^{56}$Ni mixing feature is
found with SN 2008D, SN 2009jf, and iPTF13bvn which belong to Type Ib, while
the feature of strong $^{56}$Ni mixing is found with SN 1994I which belongs to
Type Ic.  Our result also implies the possibility that  SN 2009ex that was
considered as an intermediate case (i.e., Type Ib/c) between SN Ib and SN Ic by
\citet{Hamuy2002} have relatively weak \ion{He}{1} lines because of inefficient
$^{56}$Ni mixing into the helium envelope rather than because of a small helium
content in the progenitor. On the other hand,  SN 2013ge is also classified as
Type Ib/c based on weak helium signature in the earliest spectra
\citep{Drout2016} but the signature of strong $^{56}$Ni mixing in the color
evolution of this SN and the disappearance of helium lines in later stages
provide evidence that the progenitor did not have a helium-rich envelope.  

We tentatively conclude that the population of SN Ib/Ic progenitors is largely
bimodal in terms of helium content (Section~\ref{sec:implication}).  For
further tests of our conclusions, we suggest future multi-dimensional numerical
simulations of chemical mixing using realistic SN Ib/Ic progenitor models and
more observations with detailed case studies of individual SNe Ib/Ic on the
early-time evolution. 

\acknowledgments

This work was supported by the Korea Astronomy and Space Science Institute
under the R\&D program supervised by the Ministry of Science and ICT.  SCY
thanks Alexander Heger, Francisco F{\"o}rster, Alejandro Clocchiatti, Myungshin
Im, and  Changsu Choi for useful discussions.  SCY acknowledges the support by
the Monash Center for Astrophysics via the distinguished visitor program.  AT
and SB are supported by the World Premier International Research Initiative (WPI Initiative). 
SB is supported by RSCF grant 18-12-00522 
in his work on the STELLA code development.

\bibliographystyle{aasjournal}
\bibliography{references}
\end{document}